\documentclass[a4paper,fleqn,usenatbib]{mnras}
\pdfoutput=1
\usepackage{newtxtext,newtxmath}

\usepackage[T1]{fontenc}
\usepackage{ae,aecompl}
\usepackage{times}

\usepackage{graphicx}	
\usepackage{amsmath}	
\usepackage{mathtools}
\usepackage[usenames,dvipsnames]{xcolor}

\newcommand{\cf}{cf.~}
\newcommand{\ie}{i.e.,~}
\newcommand{\eg}{e.g.,~}

\newcommand{\RU}{Department of Astrophysics/IMAPP,
  Radboud University Nijmegen, P.O. Box 9010,
  NL-6500 GL Nijmegen, the Netherlands}

\newcommand{\ITP}{Institut f{\"u}r Theoretische Physik,
  Max-von-Laue-Stra{\ss}e 1, D-60438 Frankfurt, Germany}
\newcommand{\TCD}{School of Mathematics, Trinity College, Dublin 2,
  Ireland}
  
\newcommand{\UCL}{Mullard Space Science Laboratory,
  University College London, Holmbury St.\,Mary, Dorking, Surrey RH5 6NT, UK}
  
\newcommand{\MPIfR}{Max-Planck-Institut f\"ur Radioastronomie,
  Auf dem H\"ugel 69, D-53121 Bonn, Germany}

\newcommand{\Pancini}{Dipartimento di Fisica `E.~Pancini', Universit\'a di Napoli
  `Federico II', Via Cinthia, I-80126 Napoli, Italy}

\newcommand{\INFN}{INFN Sez.~di Napoli, Via Cinthia, I-80126 Napoli, Italy}

\newcommand{\UVA}{Astronomical Institute Anton Pannekoek, University of Amsterdam,
   Science Park 904, NL-1098 XH Amsterdam, the Netherlands}

\newcommand{\JBCA}{Jodrell Bank Centre for Astrophysics,
   University of Manchester, Manchester M13 9PL, UK}

\title[How to tell a boson star from a black hole]{How to tell an accreting boson star from a black hole}

\author[H. Olivares et al.]{Hector Olivares,$^{1,6}$\thanks{E-mail: holivares@science.ru.nl}
Ziri Younsi,$^{2,6}$
Christian M. Fromm,$^{6,7}$
Mariafelicia De Laurentis,$^{3,4,6}$
\newauthor
Oliver Porth,$^{5,6}$
Yosuke Mizuno,$^{6}$
Heino Falcke,$^{1}$
Michael Kramer,$^{7,8}$ and
\newauthor
Luciano Rezzolla$^{6,9}$
\\
$^{1}$\RU \\
$^{2}$\UCL \\
$^{3}$\Pancini \\
$^{4}$\INFN \\
$^{5}$\UVA \\
$^{6}$\ITP \\
$^{7}$\MPIfR \\
$^{8}$\JBCA \\
$^{9}$\TCD \\
}

\date{Accepted XXX. Received YYY; in original form ZZZ}

\pubyear{2020}
\usepackage{hyperref}
\hypersetup{
}
\begin{document}
\label{firstpage}
\pagerange{\pageref{firstpage}--\pageref{lastpage}}
\maketitle

\begin{abstract}
The capability of the Event Horizon Telescope (EHT) to image the nearest
supermassive black hole candidates at horizon-scale resolutions offers
a novel means to study gravity in its strongest regimes and to test
different models for these objects.
Here, we study the observational appearance at 230 GHz of a surfaceless
black hole mimicker, namely a non-rotating boson star, in a scenario
consistent with the properties of the accretion flow onto Sgr A*.
To this end, we perform general relativistic magnetohydrodynamic
simulations followed by general relativistic radiative transfer
calculations in the boson star space-time. Synthetic reconstructed
images considering realistic astronomical observing conditions show that,
despite qualitative similarities, the differences in the appearance of
a black hole -- either rotating or not -- and a boson star of the type
  considered here are large enough to be detectable.
These differences arise from dynamical effects directly related to the
absence of an event horizon, in particular, the accumulation of
matter in the form of a small torus or a spheroidal cloud in the
interior of the boson star, and the absence of an evacuated
high-magnetization funnel in the polar regions.
The mechanism behind these effects is general enough to apply to other
horizonless and surfaceless black hole mimickers, strengthening
confidence in the ability of the EHT to identify such objects via radio
observations.
\end{abstract}



\section{Introduction}
\label{sec:introduction}

Observations of the Galactic Centre have confirmed the existence of a
supermassive compact object at the radio source Sgr~A*. Stellar motions
have constrained its mass to $\approx 4\times10^{6} \,{\rm M}_\odot$
\citep{Ghez:2008,Gillessen2009L,Chatzopoulos2015,Boehle2016,Abuter2018,Abuter2020}
and its density to $\approx 6\times 10^{15} \,{\rm M}_\odot\,\mathrm{pc}^{-3}$
\citep{Ghez:2008}, favouring the hypothesis of a single massive object.
Moreover, its low luminosity combined with its estimated accretion rate
indicates the absence of an emitting hard surface
\citep{Marrone2007,Broderick2009a}. All of these features are consistent
with a supermassive black hole (SMBH) as those believed to exist at the
centres of most galaxies. {Furthermore, flaring activity observed by
  the GRAVITY-Very Large Telescope Interferometer has been shown to be
  consistent with orbital motions near Sgr~A*'s last stable circular orbit
  \citep{Abuter2018b}.}
International efforts from the Event Horizon Telescope Collaboration
\citep[EHTC; ][]{Doeleman2008,Akiyama2015,Fish2016} and BlackHoleCam
\citep{Goddi2017} {successfully applied very-long-baseline interferometry
  (VLBI) techniques to obtain the first ever images of the SMBH candidate
  in the nearby galaxy M87 at a resolution comparable to the size of its
  event horizon
  \citep{Akiyama2019_L1,Akiyama2019_L2,Akiyama2019_L3,Akiyama2019_L4,Akiyama2019_L5,Akiyama2019_L6},
  and data are currently being processed to obtain analogous images for
  Sgr~A*. The M87 observations are consistent with the expectations for
  a Kerr black hole
  \citep{Akiyama2019_L1,Akiyama2019_L5,Akiyama2019_L6}}, namely, a
``crescent'' or ring-like feature, consisting of a dark region
(associated with the ``shadow'' of the black hole) obscuring the lensed
image of a bright accretion flow
\citep{Cunningham1973,Falcke2000,Grenzebach2016}. The shape of this dark
region can be exploited either to determine the properties of the black
hole within the Kerr assumption \citep{Akiyama2019_L5,Akiyama2019_L6},
or to perform tests of general relativity
\citep{Abdujabbarov2015,Psaltis2014,Younsi2016,Psaltis2016}, a
possibility assessed {for Sgr~A*} by \citet{Mizuno2018} in a realistic
scenario for the 2017 EHTC campaign and for near-future observations.

Even though the observations of the EHTC are consistent with the
image expected from an accreting Kerr black hole, it is important to
consider whether qualitatively similar images can be associated with
other kinds of compact objects, and if so, how they could be
distinguished from a Kerr black hole.
Black holes are not the only objects predicted by general
relativity that satisfy the {constraints given by the aforementioned
properties of Sgr~A*}, i.e., (1) being able to grow to millions of
solar masses, (2) being extremely compact, and (3) lacking a hard surface.
Some examples include: geons \citep{Wheeler55,Brill64,Anderson97b},
oscillatons \citep{Seidel91a,UrenaLopez2002}, Q-balls
\citep{Kleihaus2005me} and compact configurations of self-interacting
dark matter \citep{Saxton2016}. Allowing for the presence of a
surface, the list of plausible compact objects can be expanded to
include ultracompact objects with exotic surface properties, such as
gravastars \citep{Mazur2004,Cattoen2005,Chirenti2008,Chirenti2016}. 
While for black holes the photon ring plays an important role in the
formation of the shadow, it has been shown that horizonless objects
that are compact enough to produce photon rings are unstable on short
time-scales and under very general conditions, and are thus not viable as
alternatives to SMBHs \citep{Cunha2017b}. Nevertheless, there is room
for compact objects other than black holes to produce dark regions that
effectively appear as shadows, as shown for example by
\cite{Vincent:2016a} for the case of boson stars.

Boson stars are compact objects resulting from self-gravitating scalar
fields, and are a very interesting case due to the ubiquity of scalar
fields in cosmology \citep{Albrecht1982, Linde1982, Preskill1983,
  Matos1998vk, Hui2017}, string theory \citep{Arvanitaki2010}, and
extensions of general relativity such as scalar-tensor theories
\citep{Maeda2003}. Several authors have explored the possibility that
supermassive boson stars could exist at the centres of galaxies or act as
black hole mimickers \citep[see \eg][]{Schunck1997dn,
  Schunck1999, Capozziello2000, Schunck1999zu, Torres2000,
  Guzman2005, Vincent:2016a}. Consequently, a number of studies have {investigated}
the signatures of such objects, which include the dynamics of accreted
particles \citep{Schunck1999zu}, the gravitational redshift
\citep{Schunck1997dn}, and lensing \citep{Virbhadra1998,
Dabrowski2000, Virbhadra1999, Cunha2015, Cunha2017} of radiation emitted within the
boson star, and the stellar orbits around them
\citep{Grould2017}. \citet{Guzman2006, Guzman2011} studied spectra of
alpha-discs \citep{Shakura1973} around boson stars, reporting the
absence of a clear signature distinguishing them from black
holes. Motivated by the then forthcoming observations of the EHTC,
\citet{Vincent:2016a} reached similar conclusions by comparing
strong-field images of stationary tori in equilibrium around a Kerr black
hole and several boson stars. Specifically, they found that a
central dark region that mimics the shape and size of a black hole
shadow may appear for boson stars as a result of lensing of the empty
space around which the torus orbits. On the basis of this set-up, it was
concluded that boson stars would be very difficult to distinguish from
black holes by means of strong field images. While these considerations
are correct given the physical scenario considered, it is clear that
the latter does not account for the dynamics of the matter that from
the torus will accrete towards the centre of the boson star. Indeed,
uncountable astronomical observations -- and numerous numerical
simulations -- clearly indicate that quasi-stationary accretion process
accompany the dynamics of tori around compact objects.
Furthermore, the existence of stable circular orbits at all radii, at
least for spherically symmetric boson stars \citep{Guzman2006}, makes
the choice of the inner radius of the equilibrium torus arbitrary,
whereas in a realistic situation the accreted plasma is able to reach
all regions within the boson star interior.

Numerical simulations of unmagnetized zero angular momentum
accretion flows onto boson stars were carried out by
\citet{Meliani2016}, finding a significantly different behaviour
with respect to black holes as a result of the absence of an event
horizon. Specifically, they observed a polar outflow produced by
the collision of matter infalling radially from the disc. However,
this study did not include a systematic investigation of the
discernibility of the emission from the two compact objects via
ray-traced images. Moreover, accretion onto astrophysical compact objects
is believed to {occur as a result of a gradual loss of angular
momentum from orbiting matter driven by} the magnetorotational
instability \citep[MRI; ][]{Balbus1991}, and radiation at the observing
frequencies of VLBI experiments is mainly produced by synchrotron
emission. Therefore, the inclusion of magnetic fields is essential to
realistically simulate VLBI observations.

We revisit the question of the observational appearance at 230 GHz of a
boson star at the Galactic Centre, and of its distinguishability from an
SMBH. To this end, we produce strong-field synthetic EHTC images of
accreting black holes and of an accreting boson star, modelling the
accretion flow by means of fully dynamic general relativistic
ideal magnetohydrodynamic (GRMHD) simulations. Together with considering
the plasma configurations that arise from the same turbulent processes
believed to occur in nature, these simulations allow us to understand the
dynamics of accretion flows onto horizonless and surfaceless compact
objects, and to identify those features that could appear in situations
that are more general than the particular boson star case considered
here.

Using the results of these simulations, we also perform
general relativistic radiative transfer calculations and produce
synthetic images accounting for realistic EHTC observations.  As we will
highlight in what follows, we conclude that under these conditions, it is
possible to discriminate between an accreting black hole and the boson
star considered in this study by means of VLBI observations. In
particular, we show that this distinction is possible because accretion
onto the boson stars considered here leads to the accumulation of matter
down the innermost regions of the compact object. Indeed, because matter
can even reach the centre of the boson star, emission will be present at
all radii and dark regions in the image -- if they exist at all -- are
much smaller than those coming from black holes having the same mass. On
the basis of these considerations it is possible to state quite
generically that although horizonless and surfaceless objects can form
dark regions that are qualitatively similar to the shadow of a black
hole, these will be smaller than that expected size of the shadow of a
black hole of the same mass under very general circumstances.

\section{Initial data and numerical set-up}
\label{sec:physical_scenario}
We simulate numerically in three spatial dimensions (3D) the accretion
from a magnetized torus onto a Kerr black hole with total angular
momentum $J$, a Schwarzschild black hole, and two cases of non-rotating
boson stars, all with the same mass $M$. The Kerr black hole has a
dimensionless spin parameter $a:=J/M^{2}=0.9375$ (we use units with
$G=c=1$). Although results relative to the Schwarzschild black hole case
will also be presented, we first focus our discussion on the comparison
between the non-rotating boson star and the Kerr black hole.

There are two reasons behind this choice. First, rotating boson stars are
computationally more difficult to generate, requiring the solution of a
system of elliptic partial differential equations instead of the ordinary
differential equations that describe non-rotating models. Being this is
the first self-consistent study of the observational properties of
accreting boson stars, we decided to start with the simplest
configuration -- a non-rotating boson star with a simple quadratic
potential, \ie a ``mini boson star'' -- and to leave others for future
work. As will be explained below, this has lead us to results that are
applicable {to some extent to more general situations}. In addition, this
approximation might not be so severe in light of the fact that very
compact horizonless objects, including boson stars, cannot be rapid
rotators since they are subject to a dynamical instability when rotating
fast enough to produce ergoregions
\citep{Comins1978,Yoshida1996,Cardoso2007,Chirenti2008}. The more compact
the boson star, the smaller the spin parameter required to produce an
ergoregion and hence an instability. Furthermore, recent numerical
studies suggest that even slowly rotating configurations collapse either
to Kerr black holes or to non-rotating boson stars, due to a fast
instability possibly related to their topology\footnote{In the
    rotating boson star models simulated by \citet{Sanchis-Gual2019}, an
    instability develops on timescales $\lesssim10^4\,M$, where $M$ is
    the mass of the boson star. This corresponds to approximately 2.3
    days for Sgr~A*, and 10 years for M87. The study shows the existence
    of a fast instability for rotating boson stars that is not associated
    to an ergoregion, but it does not derive general instability
    conditions or timescales applicable to other cases. As a result, this
    does not prevent the existence of stable regions of the parameter
    space. Even if rotating boson stars were unstable in general, they
    still could be used as proxies for unknown solutions of the Einstein
    equations for horizonless, surfaceless, rotating objects with longer
    lifetimes, which makes future studies of their astrophysical
    appearance still relevant.}  \citep{Sanchis-Gual2019}. The existence
of this instability in the non-linear regime is supported by the fact that
rapidly rotating boson stars are not found as endpoints in the evolution
of merging binaries \citep{Bezares2017, Palenzuela2017}. Second, for
non-rotating mini boson stars, the absence of a surface or a capture cross
section permits stable circular orbits for massive particles down to the
centre of the boson star. From an observational point of view, this is
expected to lead to smaller source sizes, with emission concentrated near
the center. On the other hand, the size of a black hole image is closely
related to that of its shadow, which, in turn, is smaller for rapidly
spinning black holes. Hence, the image of a rapidly rotating black hole
will be closer in size to that of a non-rotating boson star having the
same mass, making the issue of the distinguishability much more
relevant. In addition, it is possible that the complex lensing patterns
that can be generated by rotating boson stars \citep{Cunha2015,
  Cunha2017, Vincent:2016a} would produce images that are more easily
distinguishable from those of black holes. Overall, these considerations
all suggest that interpreting strong field images is most challenging
when comparing a non-rotating boson star and a rapidly rotating black
hole.

As mentioned above, the two boson star space-times considered here
are solutions of the Einstein--Klein--Gordon system in spherical symmetry
for the potential of a mini boson star \citep{Kaup1968} (more
information on the methods used to obtain these solutions is given in
Appendix \ref{sec:background_space-time}).
For the first of these two models, which hereafter we will refer to
as ``model A'', the 99 per cent compactness is $\mathcal{C}_{99} \coloneqq
 M_{99}/R_{99} = 0.098$, where $R_{99}$ is the radius within which
$99\%$ of the mass ($M_{99}$) is contained. On the other hand, the
second model, which we will refer to as ``model B'', has a 
compactness $\mathcal{C}_{99}= 0.075$. While these compactnesses
are not the largest that can be achieved for boson stars\footnote{Quartic
potentials can achieve a higher upper limit of
$\mathcal{C}_{99}=0.16$ \citep{Amaro-Seoane2010}, while boson stars
with sextic potentials (also known as ``Q-balls'') can approach the
black hole limit of $\mathcal{C}_{99}=0.5$ \citep{Kleihaus2012}; more
complicated potentials can go arbitrarily close to it
\citep{Cardoso2016}. However, boson stars compact enough to produce
photon rings are also known to suffer from fast instabilities
\citep{Cunha2017}.}, they are among the most compact boson stars with a
quadratic potential, for which the maximum limit is $\approx 0.11$,
or $\approx 0.08$ for stable configurations.

{It is worth mentioning that although $\mathcal{C}_{99}$ is widely
  used to have a rough idea of how compact and hence ``relativistic'' a
  compact object is, for objects with large variations in density such as
  boson stars, its value could vary considerably if a different
  percentage of the mass is considered.}  {Overall, and as it will be
  shown in section \ref{sec:numerical_results} and in Appendix
  \ref{sec:mini-torus}, the two boson star models considered here are
  useful and representative cases of the two distinct behaviours of the
  accretion flow that are possible for a surfaceless compact object.}

To simulate the accretion flow we use the publicly available code
\textsc{bhac} \citep[][\url{www.bhac.science}]{Porth2017,Olivares2019},
which solves the equations of GRMHD in
arbitrary stationary space-times using state-of-the-art numerical
methods. The plasma follows an ideal-fluid equation of state with
adiabatic index $\hat{\gamma}=4/3$ \citep{Rezzolla_book:2013}. Random
perturbations are added to the initial equilibrium torus to trigger the
MRI and allow accretion. Details on the construction of the tori, and the
choices made in order to perform a fair comparison and to ensure a proper
resolution of the MRI are provided in Appendix \ref{sec:tori}. Finally,
since the mass of the accretion disc is negligible when compared to that
of the compact object (test fluid approximation), the space-time can be
considered fixed and the scalar field has no interaction with the fluid
or the electromagnetic field besides the gravitational one.

\section{Numerical results}
\label{sec:numerical_results}

\begin{figure}
\begin{center}
  \includegraphics[width=\linewidth]{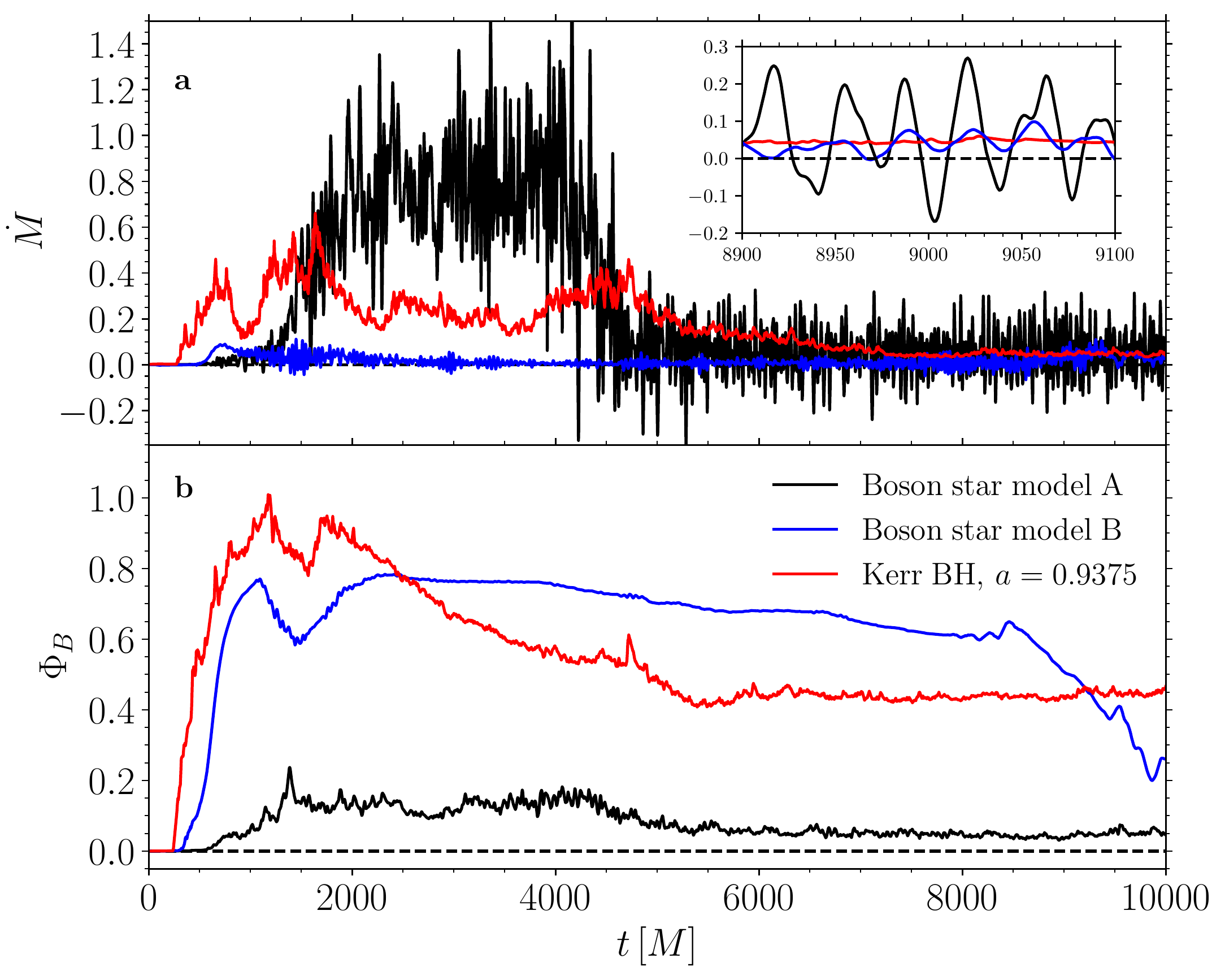}
  \caption{Evolution of the mass accretion rate ({\it a}) and the
    absolute magnetic flux ({\it b}) through the outer event horizon for
    the Kerr black hole and through a spherical shell at $r=2\,M$ for the
    two boson stars, in dimensionless units. The inset in panel
      (a) reports the mass accretion rate in the time window
      between $t=8900\ M$ and $9100\ M$. Note that mass accretion
      rate becomes quasi-stationary after $t \simeq 6000\,M$ and that
    the accretion rate can also be negative for the boson stars. 
      The drop in magnetic flux between $t/M\in [8000,10000]$ for
      boson star model B is due to a rearrangement of the internal
      magnetic field of the boson star and is discussed in more detail in
      Appendix \ref{sec:bosonB}.}
\label{fig:GRMHD_a}
\end{center}
\end{figure}

\begin{figure*}
\begin{center}
  \includegraphics[width=0.75\linewidth]{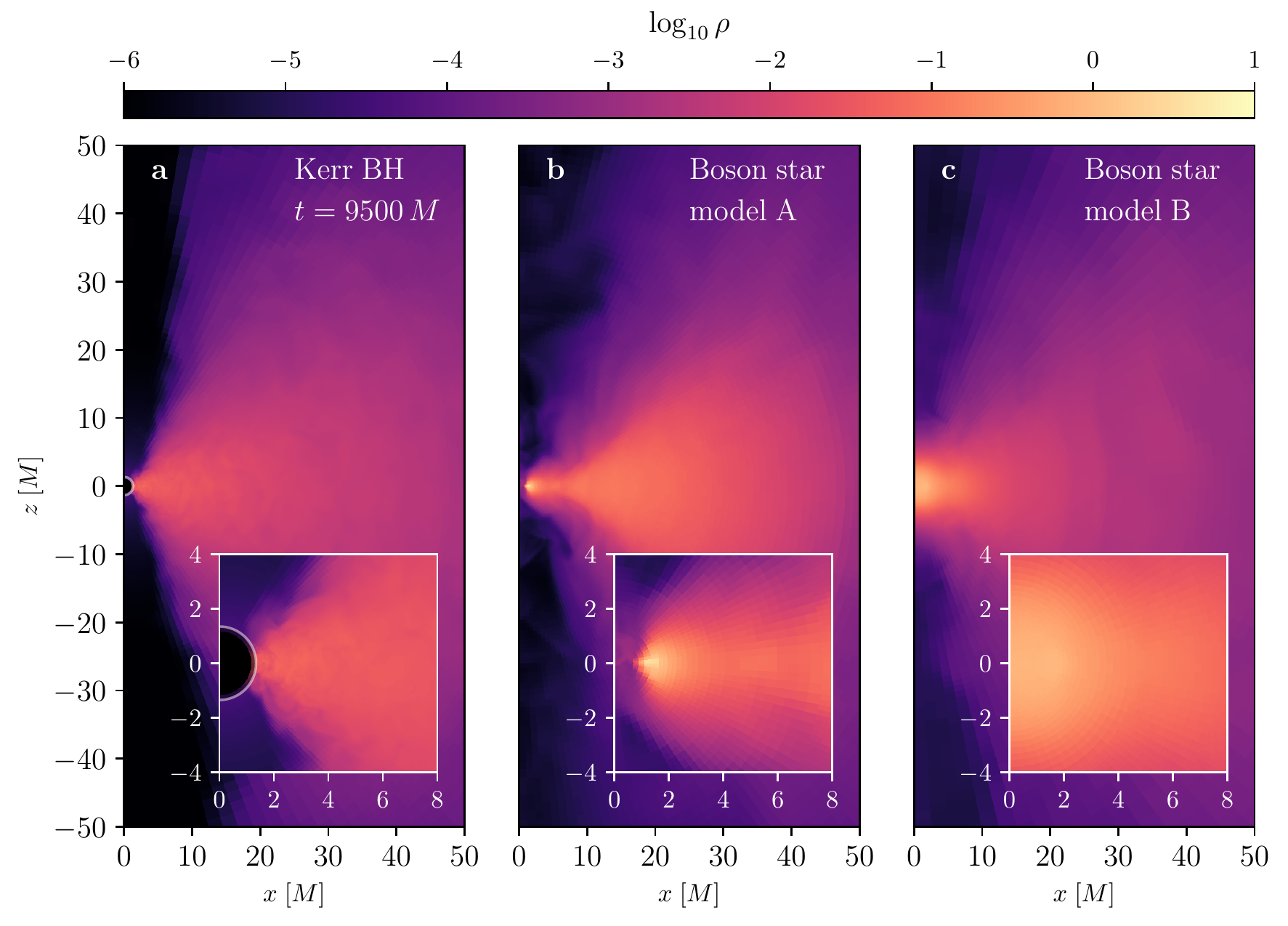}
  \includegraphics[width=0.75\linewidth]{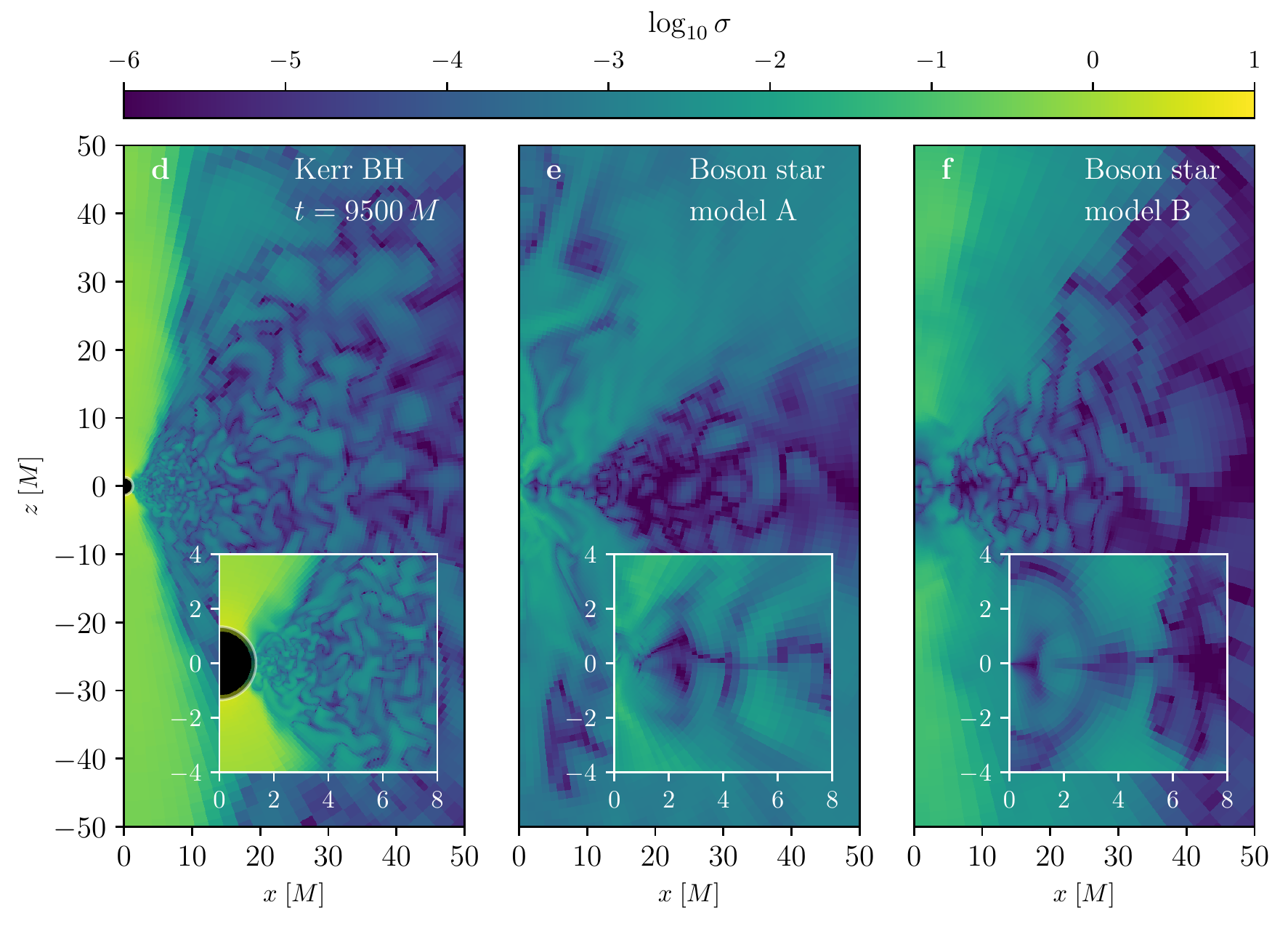}
  \caption{Rest-mass density in the fluid frame (panels a, b and c)
    and logarithmic plasma magnetization $\sigma=b^2/\rho$ (panels d,
      e and f) at $t=9500\,M$, for the Kerr black hole (a and d),
    and boson star models A (b and e) and B (c and f). The
    black hole horizon is marked by a white line and its excised interior
    is shown in solid black.}
 \label{fig:GRMHD_b}
\end{center}
\end{figure*}

As mentioned in section \ref{sec:physical_scenario}, from now on we will
focus on the comparison between the Kerr black hole case and that of the
two non-rotating boson stars.
Fig.~\ref{fig:GRMHD_a} reports in
arbitrary units the evolution of the mass accretion rate $\dot M$ (panel a)
and of the absolute magnetic flux ${\Phi}_{_{B}}$ threading a
surface at $r=r_0$ (panel b):

\begin{align}
\dot{M} &:= -\int_{r_0} \rho u^r \sqrt{-g}\, d\theta d\phi \,, \\
{\Phi}_{_{B}} &:= \frac{1}{2}\int_{r_0} |B^r|  \sqrt{-g}\, d\theta d\phi \,,
\end{align}
where $g$ is the metric determinant, $\rho$ is the rest-mass density
  of the fluid, $u^r$ is the radial component of its four-velocity, and
  $B^r$ is the radial component of the magnetic field in the Eulerian
  frame.  In the case of the black hole, we take $r_0$ to be the radial
coordinate of the outer horizon, while $r_0=2M$ for the boson stars.
After the initial growth and saturation of the MRI at $t \simeq 1000\ M$,
the mass accretion rate for each of the objects becomes
  quasi-stationary for $t \gtrsim 6000\ M$, oscillating around a small
  positive value. After $t=8000\ M$, a series of changes in the
  magnetic field structure of boson star model B reduce significantly
  the amount of magnetic flux crossing the detector shell.  Although the
  state of the magnetic field cannot be described as quasi-stationary,
  total intensity images calculated before, during and after this event
  can be still considered representative, as it is discussed in Appendix
  \ref{sec:bosonB}.  Comparing the behaviour of mass accretion rate 
  for the different objects it is possible to appreciate that while the
black hole always has a positive $\dot{M}$, a boson star can also attain
negative values. This is permitted at all radii due to the absence of an
event horizon.

As we will discuss below, this outflow is due to oscillations of an {
  internal configuration of matter accumulating within the boson star,
  whose geometric distribution can take either the shape of a mini torus
  (as in the case of model A) or of a mini cloud (for model B), depending
  on the properties of the space-time} (see Appendix \ref{sec:mini-torus}
for details). A magnified view of $\dot{M}$ during the quasi-stationary
stage of the accretion is shown in the inset of Fig.~\ref{fig:GRMHD_a}
(a), highlighting these quasi-periodic inflows and outflows.
{For the case in which the stalled accretion is in the form of a mini
  torus (model A)}, we have found the typical frequency associated with
the quasi-periodic oscillations in $\dot{M}$ to be very close to the
epicyclic frequency at the inner edge of the mini torus. This is
unsurprising since matter accumulates in this region and small
perturbations there will trigger trapped p-mode oscillations that
induce large excursions, both positive and negative, in the accretion
rate \citep{Rezzolla_qpo_03a, Rezzolla_qpo_03b}. {On the other hand,
  in the case in which the stalled accreting matter is in the form of a
  mini (spheroidal) cloud (model B), the oscillations in the accretion
  rate originate from the response of the central cloud when compressed
  by the accreting matter}.

Figure~\ref{fig:GRMHD_b} shows a snapshot at $t=9500\,M$ and on the
meridional plane, of rest-mass density $\rho$ (panels a, b and c)
and plasma magnetization $\sigma := b^2/\rho$ (panels d, e and f),
where $b$ is the magnitude of the magnetic field in the fluid frame. In
each panel we contrast the behaviour of these quantities in the case of
the Kerr black hole (panels a and d) with that of boson stars A
(panels b and e) and B (panels c and f). As anticipated, a
peculiar feature of the accretion onto the boson star of model A is the
formation of a smaller torus, which is most clearly visible in the
inset of panel (b) of Fig.~\ref{fig:GRMHD_b}.
This small torus, which essentially represents a stalled portion of the
accretion flow, is produced by the presence of both a steep centrifugal
barrier and by the suppression of the MRI. In fact, {we observe that
  for small radii}, the orbital angular velocity decreases towards the
centre, violating the criterion for the occurrence of the MRI and
stalling matter at the radius where the angular velocity profile reaches
a maximum \citep{Balbus1991}. {In Appendix \ref{sec:mini-torus}, we
  show that the formation of this structure can be related to the angular
  velocity profile of circular geodesics in the boson star space-time,
  which enables one to predict its size for other horizonless objects
  beyond mini boson stars.}

{On the other hand, in the case of the accretion onto the boson star
  of model B, this inversion in the rotation velocity profile does not
  occur, and MRI continues to drive accretion at all radii up to the
  origin, resulting in the accumulation of fluid at the centre, as can be
  seen in the inset of panel (c) in the same figure. An interesting
  question is how long it would take for these boson stars to accrete
  enough matter to form an SMBH. Although it is not
  possible to give an answer solely from a GRMHD simulation under the
  test fluid approximation, a very rough estimate will be given in
  Appendix \ref{sec:mini-torus} using the physical mass accretion rate,
  calculated in section \ref{sec:images}. }
 
As will be shown in section \ref{sec:images}, {in both of the
  boson star cases} the accumulation of matter {inside the would-be
  horizon, \ie the region of space-time with $r < 2M$,} produces an
emitting region with an intrinsic source size smaller than that expected
for a black hole. Such smaller source-sizes can be expected to be
produced under very general circumstances and would therefore provide a
signature for distinguishing surfaceless black hole mimickers. {As
  shown in Appendix \ref{sec:mini-torus}, this is the case for a large
  portion of the parameter space of mini boson stars, which includes the
  most compact and most relativistic stable configurations.} {In
  fact, although the images of model-A boson stars could be qualitatively
  similar to those of black holes, \ie by showing ring-like structures in
  some situations, the dark region will be smaller than the shadow of a
  black hole with the same mass.} {However, for model-B boson stars,
  the effective absence of such dark regions would make their images even
  more strikingly different from those of black holes}. {In general
  therefore horizon and surfaceless} compact objects are characterised
by accretion flows reaching very small radii, so that the resulting
electromagnetic emission will lead to very small source sizes and thus
very compact dark regions.

It can also be noticed that though still orders of magnitude less dense
than the rest of the simulation, the polar region in the boson star is
much less clean than that of the black hole (Figs. \ref{fig:GRMHD_b}a, b,
and c). In fact, while the black hole's
gravity is able to evacuate the polar regions and capture matter, the hot
plasma that has reached the inner regions of the boson star can become
gravitationally unbound due to its thermal energy and flow out through
the polar regions as a slowly moving wind with Lorentz factors $\Gamma
\lesssim 1.05$. This outflow, however, is of a fundamentally different
nature to that observed by \citet{Meliani2016}, which -- in a scenario
with no magnetic fields or angular momentum -- was instead caused by the
pressure increase at the stellar centre due to matter accreted radially
from the equatorial regions.

Another obvious property of the accretion flow onto our non-rotating boson
stars is the very low magnetization present along the polar regions and
that is more than two orders of magnitude smaller than in the
corresponding black hole simulations. As a result, no significant jet is
produced in both of our accreting boson star models. While this may be
the result, in part, of the choice of non-rotating models, the mass-loss
we measure is mostly due to the combination of the steep centrifugal
barrier and of the large internal energy and the magnetic energies,
rather than by a genuine MHD acceleration process, such as the one behind
the Blandford--Znajek mechanism in rotating black holes
\citep{Blandford1977}. 

On the other hand, the lack of clear signatures for the presence of
a powerful relativistic jet in Sgr A* does still allow us to consider
non-rotating boson stars as viable models to describe the compact object
at the centre of our Galaxy.  New GRMHD simulations are evidently
needed in order to determine whether relativistic jets can be produced by
rotating boson star models. We plan to investigate these scenarios in
future works.

\section{Ray-traced and synthetic images}
\label{sec:images}

We next discuss how to use the results of the GRMHD simulations to
produce ray-traced and synthetic images at the EHTC observing frequency
of $230\,{\rm GHz}$, assuming a population of relativistic thermal
electrons at temperature $T_{\rm e}$, which emit synchrotron radiation
and are also self-absorbed. Several parameters need to be fixed when
converting the dimensionless quantities evolved numerically to produce
physical images. We fix the compact object mass as $M=4.02\times
10^{6}\,{\rm M}_{\odot}\simeq0.04\,{\rm AU}$ and the distance from the source
as $7.86\,{\rm kpc}$ \citep{Boehle2016}. This sets the length
and time scalings of the general relativistic radiative transfer
calculations \citep[see \eg][]{Younsi2012, Mizuno2018} and yields the
appropriate flux scaling. Finally, we set the ion-to-electron temperature
ratio $T_{\rm i}/T_{\rm e}=3$ \citep{Moscibrodzka2009},
and choose the compact object mass accretion rate
$\dot{M}$ such that, at a resolution of $1024\times 1024$
pixels, the total integrated flux of the image reproduces Sgr~A*'s
observed flux of $\simeq 3.4\,{\rm Jy}$ at $230\,{\rm GHz}$
\citep{Marrone2006}.
{The mass accretion rates obtained after rescaling for each of the
  compact objects are displayed in Table \ref{tab:phys_mdot}.} These
values were computed as averages over the time interval $t/M\in [8900,\,
  10000]$, which, for Sgr~A*, corresponds to an observing time of $\sim
6\,{\rm h}$. At these times and over these timescales, the GRMHD
simulations have reached a state that can be considered
representative (\cf Fig.~\ref{fig:GRMHD_a} and discussion at the beginning
of Section \ref{sec:numerical_results}).

\begin{figure*}
\begin{center}
  \includegraphics[width=0.75\linewidth]{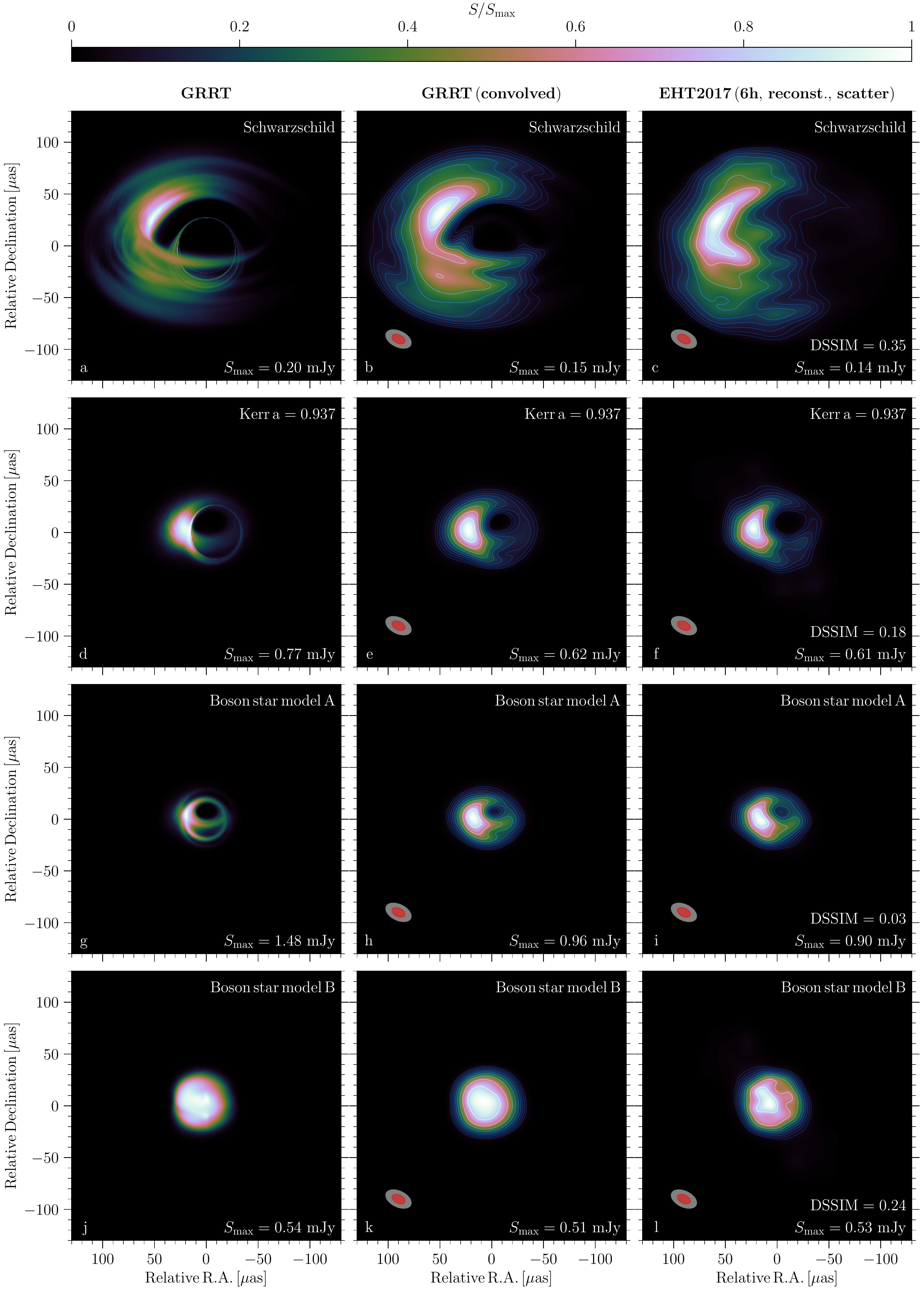}
  \caption{From top to bottom: ray-traced and synthetic images at
    $230\,{\rm GHz}$ and inclination angle of $\theta_{\rm
      obs}=60^{\circ}$ of the Schwarzschild black hole ({first row}), the
    Kerr black hole ({second row}), and boson star models A ({\it third
      row}) and B ({fourth row}). From left to right, {\it first column}:
    ray-traced images averaged over the interval $t/M\in [8900,\,
      10\ 000]$, {second column}: ray-traced images convolved with 50 per cent
    (red shaded ellipse) of the EHTC beam (grey shaded ellipse), {third
      column}: reconstructed images including interstellar scattering,
    convolved with $50\%$ (red shaded ellipse) of the EHTC beam (grey
    shaded ellipse) and indicating the value of the DSSIM metric.
    }
  \label{fig:grrt-images-inc60}
\end{center}
\end{figure*}
\begin{figure*}
\begin{center}
  \includegraphics[width=0.75\linewidth]{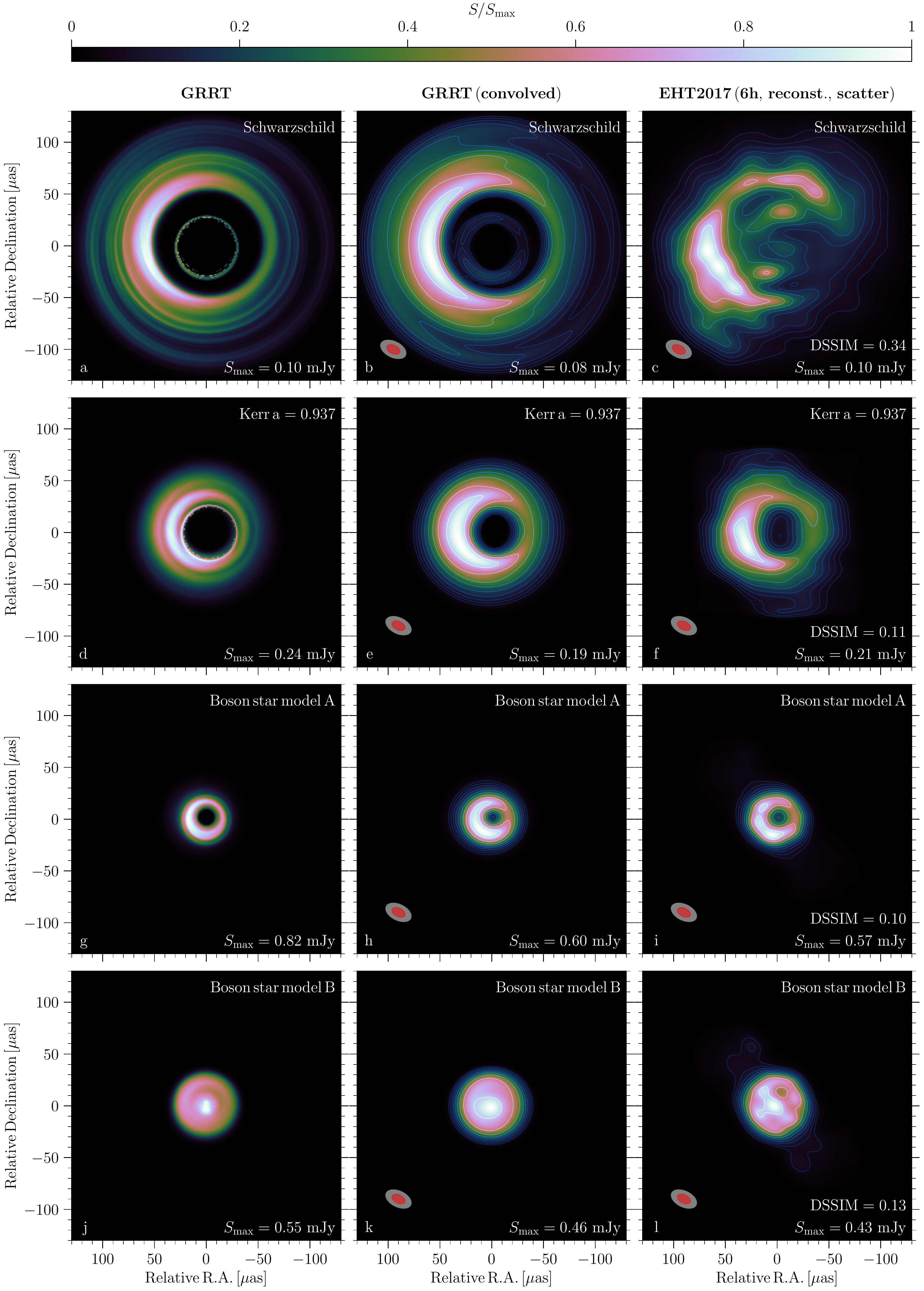}
  \caption{Same as Fig. \ref{fig:grrt-images-inc60} for an inclination
    angle of $\theta_{\rm obs}=15^{\circ}$. }
  \label{fig:grrt-images-inc15}
\end{center}
\end{figure*}

\begin{table}
        \centering
        \caption{Physical mass accretion rates (in units of $10^{-10
          }{M}_\odot\,{\rm yr}^{-1}$), obtained after rescaling the
          dimensionless accretion rates of Fig.~\ref{fig:GRMHD_a} to give
          an $\simeq 3.4\,{\rm Jy}$ flux at $230\,{\rm GHz}$ for the Kerr
          black hole and the two boson star models.}
        \label{tab:phys_mdot}
        \setlength{\tabcolsep}{0.5em}
        \begin{tabular}{lcc}
        \hline \hline
        \multicolumn{1}{|c|}{ {Object} } & {$\theta_{\rm obs}={15}^{\circ}$} & {$\theta_{\rm obs}={60}^{\circ}$} \\
        \hline
        Kerr BH        &  $34.40$  & $8.19$ \\
        BS model A  &  $8.07$    &  $6.40$  \\
        BS model B  &  $1.40$    &  $1.46$  \\
        \hline \hline
        \end{tabular}
\end{table}

In this way, using the radiative transfer code \textsc{bhoss}
\citep{Younsi2020a}, and using the same time interval mentioned above, we
produce images at several observing angles, but present here those at
$\theta_{\rm obs}=60^{\circ}$ (Fig. \ref{fig:grrt-images-inc60}),
consistent with the observational constraints found by \citep{Psaltis15},
and $\theta_{\rm obs}=15^{\circ}$ (Fig. \ref{fig:grrt-images-inc15}),
which is within the constraint $\theta_{\rm obs} \leq 27^{\circ}$ given by
hotspots models of GRAVITY observations \citep{Abuter2018b}.

We follow the same procedure to produce images for both a Kerr and a
Schwarzschild black hole. The latter is used to highlight the fact that
they differ more from those of the boson star, despite the closer
similarities of the space-time. We note, however, that the larger image
size caused by the more extended emitting region near the ISCO makes the
images produced by a Schwarzschild black hole incompatible with present
constraints on the source size of Sgr~A*, \ie $120 \pm 34\ \mu{\rm as}$
\citep{Issaoun2019}.

More specifically, the various rows of Fig.~\ref{fig:grrt-images-inc60}
show the ray-traced and synthetic images at $230\,{\rm GHz}$ and
inclination angle of $\theta_{\rm obs}=60^{\circ}$ of the Schwarzschild
black hole (first row), the Kerr black hole (second row), and boson stars
models A (third row) and B (fourth row). The different images can also be
compared across columns. From left to right, in fact, we show the average
of the ray-traced images in the interval $t/M\in [8900,\, 10\ 000]$ ({first
  column}), the same ray-traced images convolved with 50\% (red shaded
ellipse) of the EHTC beam (grey shaded ellipse; {second column}), the
reconstructed images including interstellar scattering, convolved with
$50\%$ (red shaded ellipse) of the EHTC beam (grey shaded ellipse; {third
  column}) and indicating the value of the DSSIM metric. In a very
similar fashion, Fig. \ref{fig:grrt-images-inc15} shows the equivalent
images when an inclination angle of $\theta_{\rm obs}=15^{\circ}$ is
considered.

The synthetic radio images have been generated using the \textsc{ehtim}
software package \citep{Chael2016} and after selecting as an observing
array the configuration of the EHTC 2017 observing campaign
\citep{Akiyama2019_L2}, consisting of eight radio telescopes in North
America, Europe, South America and the South Pole. To mimic realistic
radio images, we follow closely the 2017 observing schedule, using an
integration time of $12\,{\rm s}$, an on-source scan length of
$7-10\,{\rm min}$ calibration, and pointing gaps between the on-source
scans and a bandwidth of $4\,{\rm GHz}$. Within these constraints, we
perform the synthetic observations of the Galactic Center on 2017
April 8th from 08:30 to 14:30\,\textsc{UT}. The visibilities are computed by
Fourier-transforming the general relativistic radiative transfer images
and sampling them with the projected baselines of the array
\citep{Chael2016}. During this calculation, we include thermal noise and
$10\%$ gain variations, as well as interstellar scattering by a
refracting screen \citep{Johnson2015b}, as expected for the physical
condition around Sgr A*. We reconstruct the final images using a maximum
entropy method (\texttt{MEM}), provided with \texttt{ehtim}. In addition
to the calculation of the synthetic images, we convolve the
general relativistic radiative transfer images with 50\% of the EHTC beam
(second column in Fig.~\ref{fig:grrt-images-inc60}). These images can be
used to examine the influence of the sparse sampling of the Fourier space
and interstellar scattering on the reconstructed images (third column in
Fig.~\ref{fig:grrt-images-inc60}).

Overall, the visual inspection of the reconstructed images (third columns
in Figs.~\ref{fig:grrt-images-inc60} and \ref{fig:grrt-images-inc15})
shows clear differences between the four compact objects that can be
summarised as follows. First, the black hole images -- either from a
Schwarzschild or a Kerr black hole -- exhibit a ``crescent'' structure,
\ie a very asymmetric ring structure that is not present in the case of
the boson stars, whose emission tends to be either of a quasi-uniform
ring or of a uniform circle.

Second, the boson stars exhibit a smaller source size as a result of the
emission from the small torus in its interior and thus at radii
comparable or smaller than the black hole horizon. As mentioned in
section \ref{sec:numerical_results}, the location of the mini torus in
the case of model-A boson stars is determined by the radius at which the
angular velocity profile reaches a maximum. Therefore, and also for more
compact boson stars for which the exterior space-time is increasingly
similar to that of a black hole, the mini torus will be located at radii
smaller than that of the event horizon, consistently yielding a smaller
source size and a correspondingly smaller dark region as distinguishing
image features.

Third, it is possible to use the phenomenology observed in the
simulations involving boson star models A and B to calculate, in a
general way, the size of the central dark region of the class of mini
boson stars considered in this study (\cf Eq. \ref{eq:dark_size}). In
this way, we find that for all the models considered it is significantly
smaller than for black holes.  Indeed, for some modes, such as the
boson star model B, the dark region is even absent (see Appendix
\ref{sec:mini-torus} for details).  

Fourth, the boson stars generally yield a more symmetric image due to the
absence of frame dragging, which significantly reduces Doppler boosting
and consequently the sharp contrast in emission between material
approaching and receding from the observer. Given that boson stars which
are both compact and rapidly spinning are believed to be unstable, a
higher symmetry is likely to be a common property of boson star
images.

Finally, although less likely to be noticed by near-future observations
and likely requiring space-based missions \citep[see
  \eg][]{Roelofs2019}, the boson star images lack a sharp transition
between the middle dark region and its bright surroundings, which is a
fundamental property of a black hole shadow and the narrow photon
ring. In fact, due to the absence of a photon-capture cross-section, the
central dark region in the case of boson star model A is simply a lensed
image of the central low-density region.

A more quantitative assessment of the degree of similarity among the
various images considered can be made by computing image-comparison
metrics, such as the structural dissimilarity index 
\citep[DSSIM; ][]{Wang04}. The DSSIM is computed between the convolved general
relativistic radiative transfer images and the reconstructed ones and, to
guarantee that we compare similar structures within both images, we
perform an image alignment prior to its calculation and restrict to a
field of view of 110 $\mu{\rm as}$. For an inclination of $60^{\circ}$,
comparing the convolved Kerr image with the reconstructed image leads to
a DSSIM of $0.18$ and in the case of the {boson star model A} we obtain a
DSSIM of 0.03. The inter-model comparison, \ie Kerr--model A and model
A--Kerr, reveals DSSIMs of 0.31 and 0.63, respectively. Unsurprisingly,
comparisons with the Schwarzschild black hole {and with boson star model
  B} produce significantly higher DSSIM values, as reported in Tables
\ref{tab:DSSIM-inc60} and \ref{tab:DSSIM-inc15}.  Given these values, we
conclude that the models could be distinguishable with current EHTC
  observations of Sgr~A*.

  Although we plan to address this issue in more detail in a future
  work, it may be interesting to briefly discuss what are the
  consequences of our study regarding the EHT 2017 observations of M87.
  The absence of a powerful jet immediately rules out the static boson
  star models considered here as feasible models for this source.
  However, focusing only on the strong-field imaging, we may contrast the
  EHT observations with the properties of boson star images predicted by
  our simulations. Boson stars of model B, namely those for which the
  images do not display a central dark region, and which comprises all of
  those in the stable branch, are in clear contrast with the EHT
  observations, which instead show a ring-like feature. On the other
  hand, boson stars of model A produce images with ring-like structures,
  but the size of the dark region would correspond to a much larger mass
  of the central object than for the case of black holes. According to
  the estimations given in Fig. \ref{fig:b_and_rturn} (see Appendix
  \ref{sec:mini-torus-origin}), assuming the object is a boson star would
  yield a mass estimate that is $70-150\,\%$ larger than for a Kerr black
  hole, causing tension with the value obtained from stellar dynamics,
  which is in agreement with the Kerr hypothesis \citep{Akiyama2019_L1,
    Akiyama2019_L5}.

As a concluding remark we note that an additional tool to discriminate
between the two objects comes from the variability of the emission (see
Appendix \ref{sec:mini-torus} for details). Given the qualitative
differences in the accretion rate, we also expect different properties in
the energy spectra, as well as different closure-phase variabilities for
the two objects. These differences will be particularly prominent in
large antenna triangles, which probe the innermost regions currently
accessible by the EHTC.

\begin{table}
        \centering
        \caption{DSSIM metric for the comparison between the convolved
          and reconstructed images at an observer inclination angle of
          $60^{\circ}$. Self-comparisons produce significantly smaller
          values than cross-comparisons, showing that images are
          distinguishable. }
        \label{tab:DSSIM-inc60}
       \setlength{\tabcolsep}{0.5em}
        \begin{tabular}{lcccc}
        \hline \hline
          \multicolumn{1}{|c|}{ {Convolved image}} & {BH} & {BH} & {BS} & {BS}\\
       \multicolumn{1}{|c|}{} & {{(}{$\emph{a}={0}$}\bf{)}} 
       & {{(}{$\emph{a}={0.9375}$}\bf{)}} & {model A} & {model B}\\
        \hline
        BH ($a=0$)       & {\bf 0.34} &      1.03  &      0.73  &  1.04 \\
        BH ($a=0.9375$)  &      0.97  & {\bf 0.18} &      0.31  &  0.50 \\
        BS model A       &      1.21  &      0.61  & {\bf 0.03} &  0.25 \\
        BS model B       &      1.96  &      0.87  & {\bf 0.13} &  0.24 \\
        \hline \hline
        \end{tabular}
\end{table}

\begin{table}
        \centering
        \caption{Same as Table \ref{tab:DSSIM-inc60}. for an inclination
        angle of $15^{\circ}$.}
        
        \label{tab:DSSIM-inc15}
       \setlength{\tabcolsep}{0.5em}
        \begin{tabular}{lcccc}
        \hline \hline
          \multicolumn{1}{|c|}{ {Convolved image}} & {BH} & {BH} & {BS} & {BS}\\
       \multicolumn{1}{|c|}{} & {{(}{$\emph{a}={0}$}\bf{)}} 
       & {{(}{$\emph{a}={0.9375}$}\bf{)}} & {model A} & {model B}\\
        \hline
        BH ($a=0$)       & {\bf 0.34} &      0.82  &      1.22  &  1.01 \\
        BH ($a=0.9375$)  &      0.87  & {\bf 0.10} &      0.34  &  0.12 \\
        BS model A       &      1.16  &      0.26  & {\bf 0.10} &  0.28 \\
        BS model B       &      1.12  &      0.38  &      0.14  &  {\bf 0.13} \\
        \hline \hline
        \end{tabular}
\end{table}

\section{Conclusions}
\label{sec:conclusion}

We have carried out the first 3D GRMHD simulations of disc accretion onto
boson stars and combined them with general relativistic
radiative transfer calculations, with the goal of determining whether,
{under realistic observing conditions such as those of the EHTC}, an
accreting non-rotating boson star can be distinguished from a black hole
of the same mass. For the latter, we have considered both non-rotating and
rotating black holes, focusing on the second ones as they provide more
images that are more compact and hence closer to those produced by boson
stars.

By comparing the images produced for the two compact objects using very
similar set-ups, we found important differences, both in the plasma
dynamics and in the general relativistic radiative transfer
images. Indeed, the absence of a capturing surface in the case of boson
stars, introduces important and fundamental differences in the flow
dynamics. More specifically, matter accreting onto the boson stars can
reach their innermost regions, attaining quasi-stationary configurations
with either distributions that are either toroidal (\ie a mini torus) or
quasi-spheroidal (\ie a mini cloud). This behaviour, which has not been
reported before, is simply the result of the existence of stable orbits
at all radii and to the suppression of the accretion process due to
the suppression of the MRI and to the presence of a steep centrifugal
barrier. In turn, this matter behaviour leads to the absence of an
evacuated high-magnetization funnel in the polar regions and to images
that show a markedly smaller source size and a more symmetric emission
structure, in stark contrast to the characteristic crescent of the images
resulting from the accretion onto black holes. As a result of these
differences in the plasma dynamics and emission, we conclude that it is
possible to distinguish the images of the accreting mini boson star
models considered here from the corresponding images of accreting black
holes having the same mass.

The results presented have been obtained for two representative cases
of mini boson stars that are non-rotating and do not have a photon
orbit. While other boson star models could be investigated -- for
instance, by considering more complex potentials leading to more compact
solutions and even to the appearance of an unstable photon orbit -- we
believe that the results found here will continue to apply and be a
generic property also as for other surfaceless and horizonless compact
objects. This rationale is based on three important properties shared by
these objects. First, horizonless and surfaceless objects permit the
accumulation of matter within their interior. For monotonically
decreasing angular velocity profiles, this accumulation will occur at the
centre, while for angular velocity profiles having a maximum, this will
occur at this maximum in the form of a stalled mini torus.
As discussed in Appendix \ref{sec:mini-torus-origin}, for very
compact objects that have exterior space-times similar to those of black
holes, this feature will generally occur at radii smaller than that
of the event horizon of the corresponding black hole space-time, inevitably
resulting in a smaller observed image size. Second, because horizonless compact
objects rotating sufficiently fast to produce ergospheres are unstable,
the asymmetry produced by Doppler boosting {and related to the frame
  dragging} in black hole images is likely to be less pronounced for
horizonless objects. Finally, the central dark region that can be
produced by these objects does not result from a photon capture
cross-section as is the case for a black hole. Rather, it represents the lensed
image of the central low-density region, which has a diffused
boundary. As a result, the corresponding shadow can be expected to have a
much reduced brightness contrast and a sharper edge, which can be
properly revealed by imaging at increased resolutions.
All of these considerations need to be corroborated by additional
simulations, which we plan to perform in the near future. In particular,
it would be very interesting to verify whether the complex lensing
patterns produced by rotating boson stars -- as those found by
\citet{Vincent:2016a} and \citet{Cunha2017} -- do indeed facilitate
distinguishing them from black holes, when produced in a realistic
observational scenario.

Finally, we note that ongoing pulsar searches around Sgr A*
\citep{Kramer2004}, when successful, could provide additional important
information to the experiment outlined here. A suitable pulsar orbiting a
rotating boson star would enable a precise determination of its spin and
possibly even its quadrupole moment, providing valuable input for
interpretation of the image and complementary tests \citep{Wex1998,
  Liu2012, Psaltis2016}. Details on this will be part of future
work. Overall, our results and the ability to distinguish between these
compact objects underline the potential of EHTC observations to extend
our understanding of gravity in its strongest regimes and to potentially
probe the existence of self-gravitating scalar fields in astrophysical
scenarios.

\section*{Acknowledgements}
We thank T. Bronzwaer, A. Cruz-Osorio, J. Davelaar, A. Grenzebach,
D. Kling, J. K\"ohler, T. Lemmens, E. Most, M. Mart\'inez Montero,
H.-Y. Pu, L.  Shao, B. Vercnocke, F. Vincent, N. Wex, and M. Wielgus
for useful input. Support comes from the ERC Synergy Grant
``BlackHoleCam -- Imaging the Event Horizon of Black Holes'' (Grant
610058), the LOEWE-Program in HIC for FAIR. HO was supported in part by a
CONACYT-DAAD scholarship, and a Virtual Institute of Accretion (VIA)
postdoctoral fellowship from the Netherlands Research School for
Astronomy (NOVA). ZY is supported by a Leverhulme Trust Early
Career Fellowship and acknowledges support from the Alexander von Humboldt
Foundation. The simulations were performed on the SuperMUC cluster at the
Leibniz Supercomputing Centre (LRZ) in Garching, and on the LOEWE and
Iboga clusters in Frankfurt.
This work made use of the following software libraries not cited in the
text: \textsc{matplotlib} \citep{Hunter2007}, \textsc{numpy} \citep{Oliphant2006}.
This research has made use of NASA's Astrophysics Data System.

\section*{Data availability}
The data underlying this article will be shared on reasonable request
to the corresponding author.








\appendix

\section{The boson star space-time}
\label{sec:background_space-time}

As mentioned in section \ref{sec:physical_scenario}, to obtain the boson
star space-time we solve in spherical symmetry the Einstein--Klein--Gordon
system of equations for a complex scalar field $\Phi$ with the potential
of a mini boson star \citep{Kaup1968}
\begin{equation}
\label{eq:mini_bs_potential}
V(|\Phi|)= \frac{1}{2} \frac{m^2}{M_{_{\mathrm{Pl}}}^4} |\Phi|^2 \,,
\end{equation}
\noindent
where $M_{_{\mathrm{Pl}}}$ is the Planck mass. The method for computing
these configurations is presented in a number of works \citep[see
  \eg][]{Kaup1968, Ruffini1969, Liebling2012}. In brief, we start from the
Ansatz
\begin{equation}
\label{eq:phi_ansatz}
\Phi=\phi (r) e^{-i\omega t}\,,
\end{equation}
\noindent
for the scalar field, and
\begin{equation}
\label{eq:metric_ansatz}
ds^2 = -\alpha^2 dt^2 + \gamma_{rr} dr^2 + r^2d\Omega^2\,,
\end{equation}
for the metric, where $\phi$, $\alpha$ and $\gamma_{rr}$ are real
functions of the radial coordinate $r$ only. The line element in
equation (\ref{eq:metric_ansatz}) is a special case that follows from the
general 3+1 metric
\begin{equation}
\label{eq:3+1metric}
g_{\mu\nu} = \gamma_{\mu\nu} - n_\mu n_\nu \,,
\end{equation}
when the four-velocity of Eulerian observers $n^\mu =
(1/\alpha,-\beta^i/\alpha)$ has zero shift ($\beta^i=0$), and after a
particular choice of spherical coordinates \citep[see
][]{Rezzolla_book:2013}.

\begin{figure}
\includegraphics[width=\linewidth]{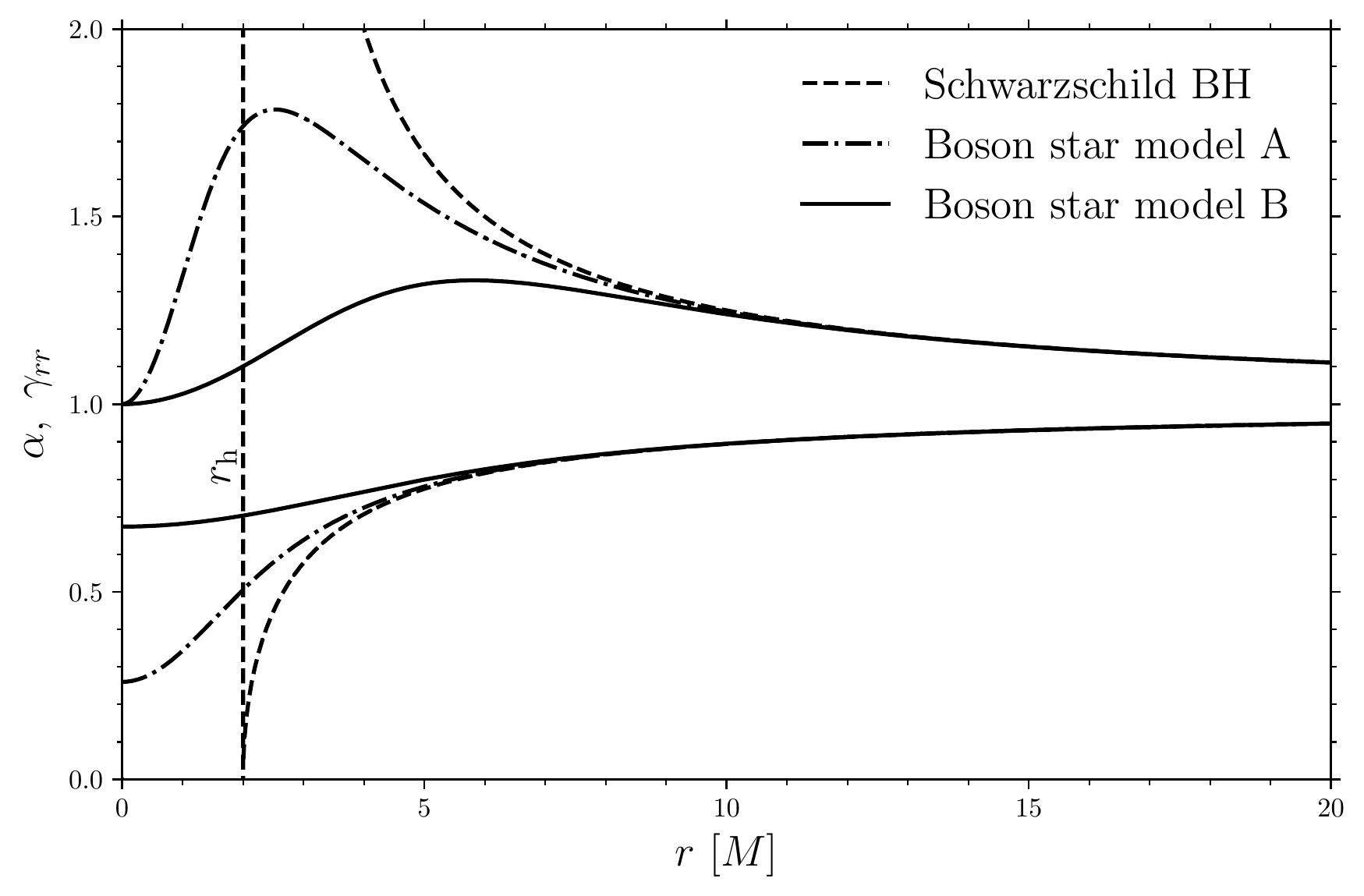}
\caption{Comparison between the metric functions of the boson star models
  used in this work and those of a Schwarzschild black hole in
  Boyer--Lindquist coordinates. The vertical dashed line shows the
  position of the black hole event horizon.}
\label{fig:metric_comparison}
\end{figure}

Upon substitution of Eqs.~(\ref{eq:phi_ansatz}) and
(\ref{eq:metric_ansatz}) in the Einstein--Klein--Gordon system, we obtain a
system of four ordinary differential equations, which we integrate by
means of the fourth-order Runge--Kutta method, enforcing asymptotic
flatness with a shooting method. { Of the models considered here,
  boson star model A has an oscillation frequency $\omega \,M \approx
0.32$ and a scalar particle mass of $m \approx 0.410 \,
(M_{_{\mathrm{Pl}}}/M) M_{_{\mathrm{Pl}}}$, while boson star model B
  has an oscillation frequency $\omega \,M \approx 0.54$ and a
  scalar particle mass of $m \approx 0.632 \, (M_{_{\mathrm{Pl}}}/M)
M_{_{\mathrm{Pl}}}$}. A comparison between their metric functions and
those of a Schwarzschild black hole is shown in
Fig. \ref{fig:metric_comparison}. For the measured mass of Sgr~A*, $M
\simeq 4.02 \times 10^6 \, {\rm M}_{\odot}$ \citep{Boehle2016}, {both
  cases} correspond to $m\approx10^{-17} \, \mathrm{eV}/c^2$, which is
within the range allowed by astronomical observations
\citep{Amaro-Seoane2010}.


  If parametrized by the central amplitude of the scalar field, the
  parameter space of mini boson stars consists of a stable and an
  unstable branch, which are separated by the maximum possible mass, $M
  \approx 0.633 \, (M_{_{\mathrm{Pl}}}/m) M_{_{\mathrm{Pl}}}$ \citep[see
    \eg ][]{Amaro-Seoane2010}. A larger amplitude is
  associated with a higher gravitational redshift, and therefore boson
  stars on the unstable branch might be considered more relativistic than
  those on the stable one, despite not possessing a higher compactness in
  the traditional sense. Boson star model A sits on the unstable branch,
  while boson star model B is on the stable branch.  Numerical
    simulations \citep{Seidel90b,Guzman2004} show that perturbed boson
    stars in the unstable branch either collapse into black holes or
    decay to lower mass stable boson stars in a time-scale of a few tens
    of oscillation periods, which for boson star model A corresponds to
    less than one hour for Sgr~A* and nearly a month for M87. Despite
  these differences, the use of the two models considered here is made
  independently of their stability properties and only with the goal of
  exploring the two possible behaviours of the accretion flow that can
  take place for a horizonless and surfaceless compact object, and that
  would lead to the formation of either a mini torus or a mini cloud at
  the boson star centre.

{As discussed in more detail in Appendix \ref{sec:mini-torus}, we
  find that these different behaviours depend in a simple way on the
  space-time properties, and therefore it is possible to predict what kind
  of accretion flow will appear in other such objects besides mini boson
  stars. In this sense, it is possible that the behavior of the accretion
  flow that we observe here for the unstable boson star (\ie the
  formation of the mini torus) may appear in horizonless and surfaceless
  compact objects that are stable.}

\section{Plasma dynamics in the boson star interior}
\label{sec:mini-torus}

\begin{figure}
\centering \includegraphics[width=\linewidth]{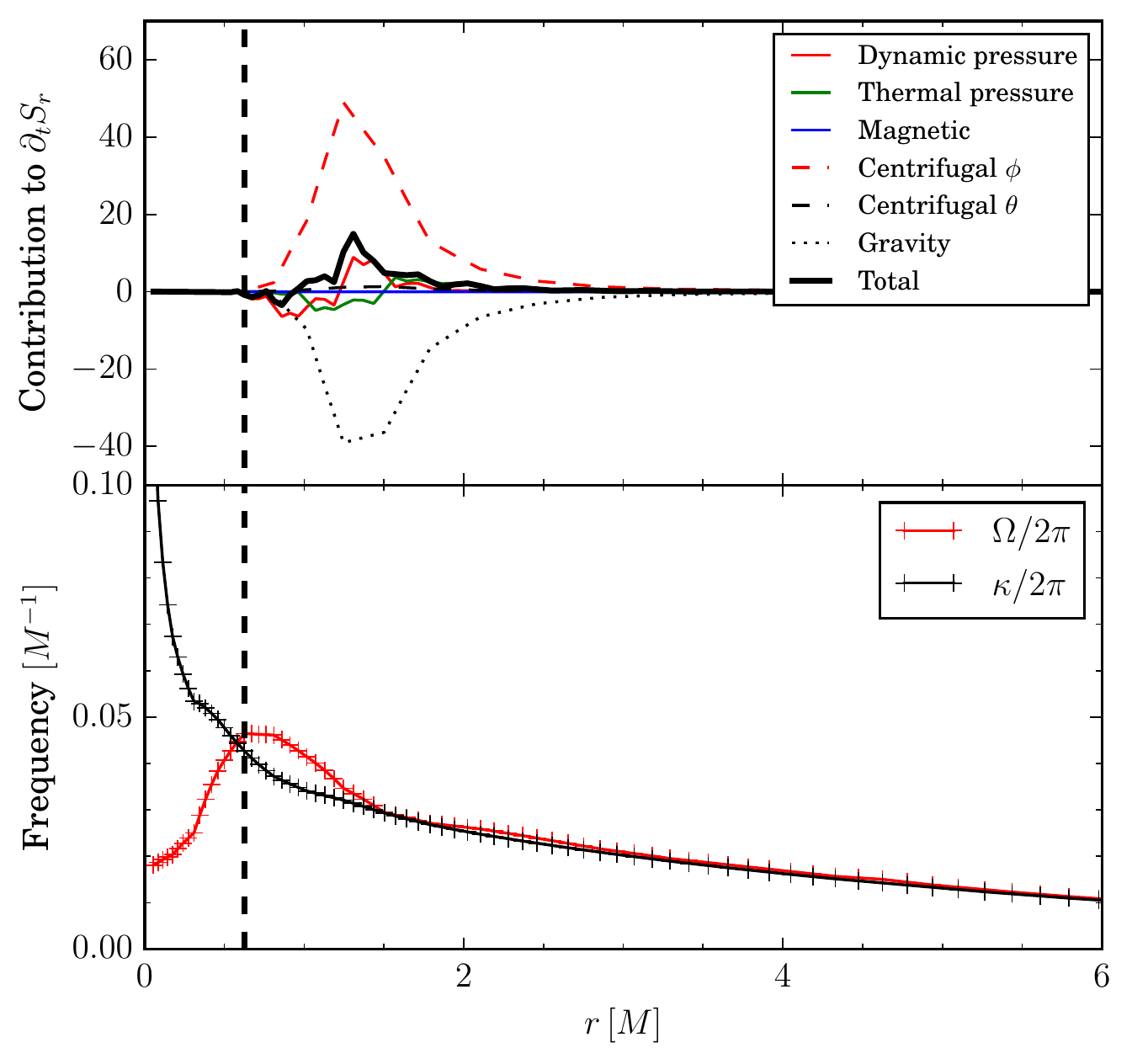}
\caption{{\it Top:} Different contributions to the conservation equation
  of radial momentum (see Eq. \ref{eq:dSr_split}) for the accretion
  flow onto boson star model A. {\it Bottom:} Orbital
  ($\Omega$) and radial epicyclic ($\kappa$) frequencies of the fluid in the
  boson star interior. Both plots consider time and $\phi$-averages of
  quantities at the equatorial plane, over the interval $t=8900 -
  10\ 000\ M$. The vertical dashed line marks the position of the turning
  point of the angular velocity, $r_{\rm turn}$, so that on the left of
  the dashed line the flow is stable to the MRI, while on the right it is
  MRI unstable.}
\label{fig:forces_rot_profiles}
\end{figure}


\subsection{Origin of the stalled mini torus}
\label{sec:mini-torus-origin}

Without an event horizon or a hard surface, a boson star also lacks a
capture cross-section. As a consequence, steep centrifugal barriers
appear for all angular momenta (except exactly zero) and it is possible
to find stable circular orbits at all radii. Indeed, as discussed in the
main text, our simulation of accretion onto boson star model A lead to
the formation of a ``hole'', that is, a spatial region at the centre of
the boson star with very low density material and surrounded by a dense
accumulation of matter in a toroidal distribution, \ie a mini torus.

To investigate the origin of this feature, we recall that the plasma
obeys the equations for local conservation of rest mass, energy, and
momentum
\begin{align}
&\nabla_{\mu}\left( \rho u^{\mu}\right) = 0 \,,  \\
&\nabla_{\mu} T^{{\mu \nu}} = 0 \,, 
\end{align}
where $\nabla_{\mu}$ denotes the covariant derivative, and $T^{{\mu
    \nu}}$ is the energy--momentum tensor of the fluid and the magnetic
field
\begin{equation}
T^{\mu\nu} = \left(\rho h + b^2 \right) u^{\mu}u^{\nu} + \left( p + b^2/2
\right) g^{\mu \nu} -b^{\mu}b^{\nu}\,. \label{eq:Tmunu}
\end{equation}
Here, $\rho$ is the rest-mass density, $h$ the fluid specific enthalpy,
$p$ the thermal pressure and $b^\mu$ the components of the magnetic
field, all measured in the fluid frame \citep[see ][]{Porth2017}. After
adopting the $3+1$ decomposition of the space-time described by equation
(\ref{eq:3+1metric}), it is possible to obtain an evolution equation for
each component of the covariant three-momentum $S_i \coloneqq
\gamma_i^{\ \mu}n^{\nu} T_{\mu\nu} $. Since accretion is best captured by
the conservation of radial momentum, it is useful to group the various
terms appearing in the conservation equation of $S_r$ and to associate with
each term the corresponding physical origin. More specifically, after
assuming symmetry in the $\phi$ direction and with respect to the
equatorial plane, the different contributions to the evolution of $S_r$
can be listed as
\begin{align}
\label{eq:dSr_split}
\partial_t S_r &= & \\
 \text{Thermal pressure:}   &-\partial_r \sqrt{\gamma} \alpha p & \nonumber \\
 \text{Dynamic pressure:}   &-\partial_r \sqrt{\gamma} (\alpha v^r - \beta^r)\rho h \Gamma^2 v_r  & \nonumber \\
 \text{Magnetic forces:}   &-\partial_r \sqrt{\gamma} \{ &  \nonumber \\
                 & (\alpha v^r - \beta^r)[\mathbf{B}^2 v_r - (B^j v_j) B_r ] & \nonumber \\
                 &- \alpha B^r [(B^j v_j) v_r + B_r/\Gamma^2] & \nonumber \\
                 & +\alpha b^2 /2 \} & \nonumber \\
 \text{Centrifugal in }\theta: &+ \sqrt{\gamma} \frac{1}{2} \alpha W^{\theta\theta}\partial_r \gamma_{\theta\theta} & \nonumber \\
 \text{Centrifugal in }\phi:   &+ \sqrt{\gamma} \frac{1}{2} \alpha W^{\phi\phi}\partial_r \gamma_{\phi\phi} & \nonumber \\
 \text{Shift:}              &+ \sqrt{\gamma} S_i \partial_r \beta^i & \nonumber \\
 \text{Gravity:}            &+\frac{1}{2} \alpha W^{ik} \partial_r \gamma_{ik} - U \partial_r \alpha & \nonumber \\
                 & - W^{\theta\theta}\partial_r \gamma_{\theta\theta} - W^{\phi\phi}\partial_r \gamma_{\phi\phi}  \nonumber\,,
\end{align}
where $\sqrt{\gamma}$ is the square root of the three-metric determinant,
$B^i$ and $v^i$ are the components of the magnetic field and the fluid
three-velocity, $W_{ij} \coloneqq \gamma_{i\mu}\gamma_{j\nu}T^{\mu\nu}$
those of the covariant stress tensor and $U := n_\mu n_\nu T^{\mu\nu}$
the total energy density, all defined in the Eulerian frame. In Eq.
(\ref{eq:dSr_split}), both magnetic pressure and tension are considered
under the label ``magnetic forces``.

The upper panel of Fig. \ref{fig:forces_rot_profiles}, reports the
numerical values of the various contributions to the conservation
equation of radial momentum in Eq. \ref{eq:dSr_split} after averaging in
time and in the $\phi$-direction. Comparing these contributions it
becomes clear that the dominant term balancing gravity is the centrifugal
force in $\phi$, while the evolution of radial momentum towards the
equilibrium state is guided by dynamic pressure. The contribution
labelled as ``shift'', which results from the movement of Eulerian
observers with respect to the coordinate system, is zero for the case
considered here and is therefore omitted in
Fig. \ref{fig:forces_rot_profiles}.

The bottom panel of Fig. \ref{fig:forces_rot_profiles} shows instead the
orbital ($\Omega$) and radial epicyclic ($\kappa$) frequencies -- after averages
in time and $\phi$-direction -- of the fluid in the boson star
interior. Note that while the orbital frequency is monotonically
decreasing outwards in the outer parts of the flow, where it follows an
essentially Keplerian fall-off, it also exhibits a local maximum and a
decreasing branch as it tends to $r \to 0$. This behaviour is due to the
decrease in the gravitational forces in the innermost regions of the
boson star and hence to a decrease in the angular momentum needed to
maintain a circular orbit.
%
%
As a result, the stability criterion against the MRI, which is given by
$d\Omega^2/dR>0$, where $R:=r\sin \theta$ \citep{Balbus1991}, is
fulfilled in the innermost regions of the boson star, where the MRI is
essentially quenched. Under these conditions, the matter in the mini
torus is unable to lose angular momentum and will be repelled by
the centrifugal barrier at the radius where $d\Omega^2/dR=0$ and forced
to move along the polar directions, where the fluid density is lower.
The bottom panel of Fig. \ref{fig:forces_rot_profiles} also shows that
this radial location coincides with the inner edge of the torus in the
equatorial plane.

{It is interesting to note that the conditions discussed above for the
  formation of the stalled torus are not met for all mini boson
  stars. Indeed, for a large part of the parameter space, which includes
  the most compact, or more relativistic, stable configurations such as
  the boson star model B, the rotation velocity profile of circular
  geodesics has no local maxima for $r>0$. As a result, the MRI is active
  at all radii and the plasma continues accreting down to the centre of
  the boson star.}

\begin{figure}
\centering
\includegraphics[width=\linewidth]{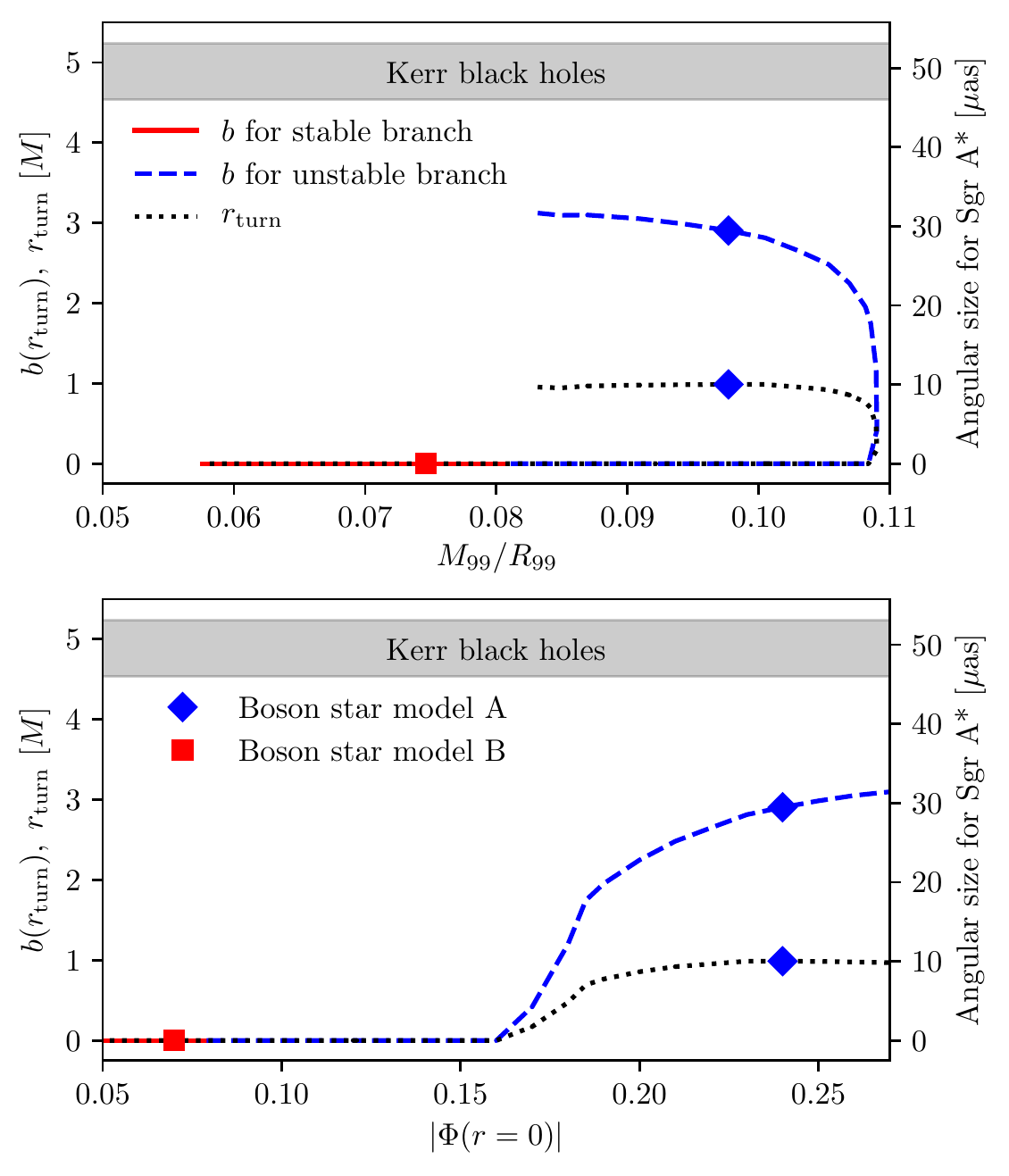}
\caption{Radial position $r_{\rm turn}$ at which the MRI is suppressed
    (dotted line), and the impact parameter $b$ for photons at this
    radius (dashed and continuous lines) for different mini boson stars,
    as a function of compactness ({\it top panel}) and central amplitudes
    of the scalar field ({\it bottom panel}). As a reference, the shadowed
    gray region shows the possible minimal widths for Kerr black hole
    shadows. Note that all boson star models of the type
    considered here have dark regions that are smaller than those associated
    with black holes. The right axis shows the corresponding size of
    the dark region associated with $b$ in $\mu$as and for the case of
    Sgr~A*. The dashed blue line corresponds to the unstable branch and
    the red continuous line to the stable branch of the mini boson star
    family. The boson star models considered in this work are indicated
    by markers. }
\label{fig:b_and_rturn}
\end{figure}

Computing the angular velocity corresponding to a circular time-like geodesic
for a massive particle as $\Omega := u^{\phi} / u^t = [ (\alpha/r)\, d\alpha/dr
]^{1/2}$ \citep[see \eg][]{Rezzolla_book:2013}, we can estimate the
location of the edge of the mini torus with the corresponding turning
point $r_{\rm turn}$ in the two branches for $r\to 0$ and $r\to
\infty$\footnote{In reality, the motion at the inner edge of the mini
  torus is expected to be non-Keplerian, but as shown in the bottom panel
  of Fig. \ref{fig:forces_rot_profiles}, $r_{\rm turn}$ is expected to
  provide a rather accurate approximation.}. Similarly, we can compute
the corresponding photon impact parameter at $r_{\rm turn}$ as
\begin{equation}
\label{eq:dark_size}
  b\,(r_{\rm turn}) = \frac{r_{\rm turn}}{\alpha\,(r_{\rm turn})}\,,
\end{equation}
and use $b\,(r_{\rm turn})$ to estimate the radial size of the ``dark
region'' in an accreting boson star of model A. Figure
\ref{fig:b_and_rturn} shows the radius $r_{\rm turn}$ and the impact
parameter $b$ for photons reaching this radius (dashed and continuous
lines) for different mini boson stars, as a function of compactness ({\it
  top panel}) and central amplitudes of the scalar field ({\it bottom
  panel}). As a reference, a shadowed gray region shows the possible
minimal widths for a Kerr black hole shadow, from $a=0$ to $a=1$. Also
as a reference, the right axis shows the corresponding size of the dark
region associated with $b$ in $\mu$as and for the case of Sgr~A*. The
dashed blue line corresponds to the unstable branch and the red
continuous line to the stable branch of the boson star family, with the
markers indicating the boson star models considered here. Overall, Fig.
\ref{fig:b_and_rturn} underlines that while strong field images of boson
stars with $r_{\rm turn}=0$, and hence with no central dark region,
are obviously going to be drastically different from those of black holes,
none of the boson stars considered here produces a dark region with size
comparable to that of the black hole shadow with the same mass.

  An interesting question is how general this property is amongst
  surfaceless and horizonless black-hole mimickers. In the discussion
  above, we showed that a necessary condition for the formation of the
  stalled mini torus, and hence of a central dark region, is the
  existence of a maximum in the angular velocity profile of the fluid,
  which -- after the re-distribution of angular momentum by turbulence --
  follows approximately that of time-like equatorial circular
  geodesics. Black-hole space--times do not have maxima in such rotation
  profiles outside the event horizon; therefore, if the exterior
  space-time of the black hole mimicker is similar to that of a black
  hole, any maximum should occur in the interior of the object. For very
  compact objects with most of their mass-energy enclosed in a radius
  comparable to their Schwarzschild radius, the inner edge of the mini
  torus would then be located at an even smaller radius. In the case of
  slowly rotating compact objects, the (Jebsen)-Birkhoff theorem makes
  the above reasoning particularly relevant \citep{Rezzolla_book:2013}.

\subsection{Quasi-periodic oscillations}

\begin{figure}
\centering
\includegraphics[width=\linewidth]{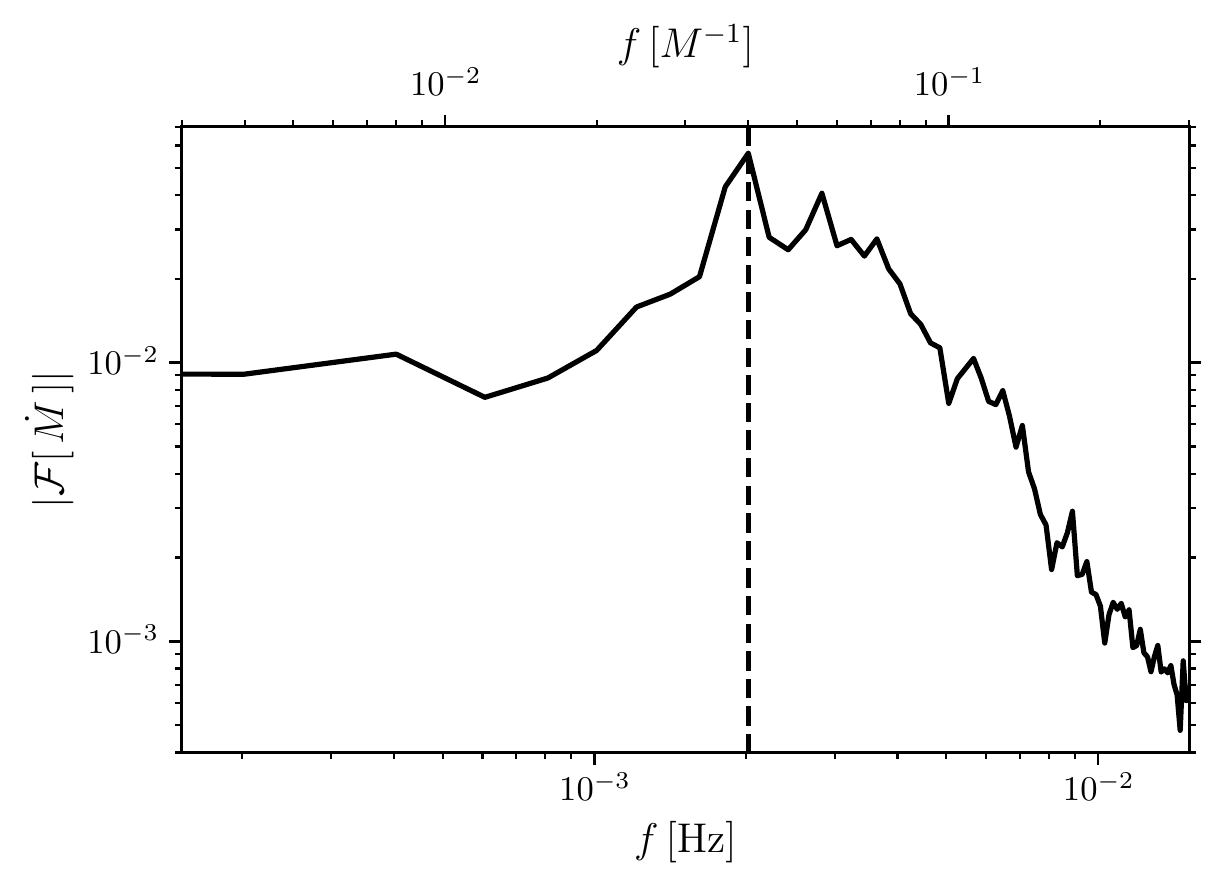}
\caption{Power spectral density of the mass accretion rate at $r=2\,M$ for
  boson star model A. A peak can be observed at $f=0.002\ {\rm Hz}$,
  which corresponds to the radial epicyclic frequency at the inner edge
  of the mini torus (black dashed line).}
\label{fig:fourierboson}
\end{figure}

As anticipated in section \ref{sec:numerical_results}, another
peculiarity of accretion onto the {boson stars} is the presence of strong
quasi-periodic oscillations in the mass inflow. {It has been shown that
  for the case of black holes accreting at rates similar to those of
  Sgr~A* and M87, the time series of the accretion rate can be used as a
  proxy to study the variability at the typical observing frequencies of
  the EHT \citep{Porth2017}}. By calculating the power spectral density
(PSD) of the these time series (Fig. \ref{fig:fourierboson}), it can be
observed that {for the case of boson star model A} the frequency peaks
around $f\approx 0.04\,M^{-1} = 0.002\ {\rm Hz}$, which closely
corresponds to the radial epicyclic frequency $\kappa/2\pi$ at the location of
the inner edge of the torus (\cf Fig. \ref{fig:forces_rot_profiles}).
The PSD reported in Fig. \ref{fig:fourierboson} was obtained by averaging
that of $10$ not overlapping time windows in the interval $5000 -
10\ 000\ M$. The large amplitude of these oscillations is caused by the
high density in the mini torus, which results in the displacement of a
large amount of mass with every cycle. As mentioned in the main text,
QPOs near the epicyclic frequency are expected from trapped p-mode
oscillations that induce large excursions, both positive and negative, in
the accretion rate \citep{Rezzolla_qpo_03b,Rezzolla_qpo_03a}.

Hence, a possible detection of QPOs in the mass accretion rate could
provide additional means for distinguishing accreting black holes from
boson stars, as we could expect the latter to show quasi-periodic
oscillations at higher frequencies. In fact, for circular orbits around
black holes, the epicyclic frequency decreases to zero at the innermost
stable circular orbit and becomes imaginary closer to the black hole
\citep{Kato1980, Abramowicz2003b}.

\subsection{Variability in the images of boson star model B}
\label{sec:bosonB}

\begin{figure}
	\centering
	\includegraphics[width=\linewidth]{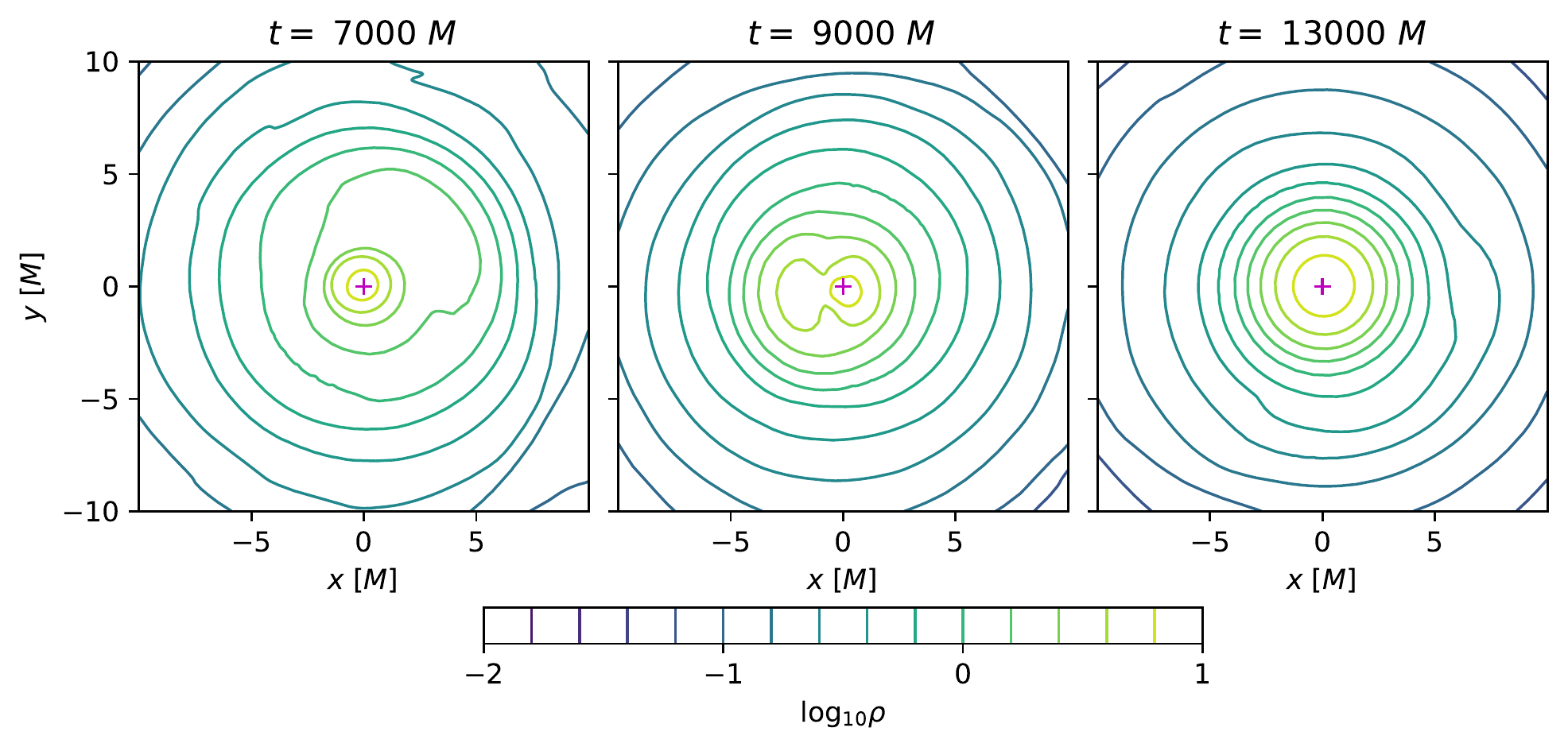}
	\caption{Isocontours of the rest-mass density for boson star
          model B at different times before, during, and after the
          absorption of the cloud discussed in Appendix
          \ref{sec:bosonB}. The red cross marks the centre of the boson
          star.}
	\label{fig:contours}
\end{figure}

\begin{figure}
	\centering
	\includegraphics[width=\linewidth]{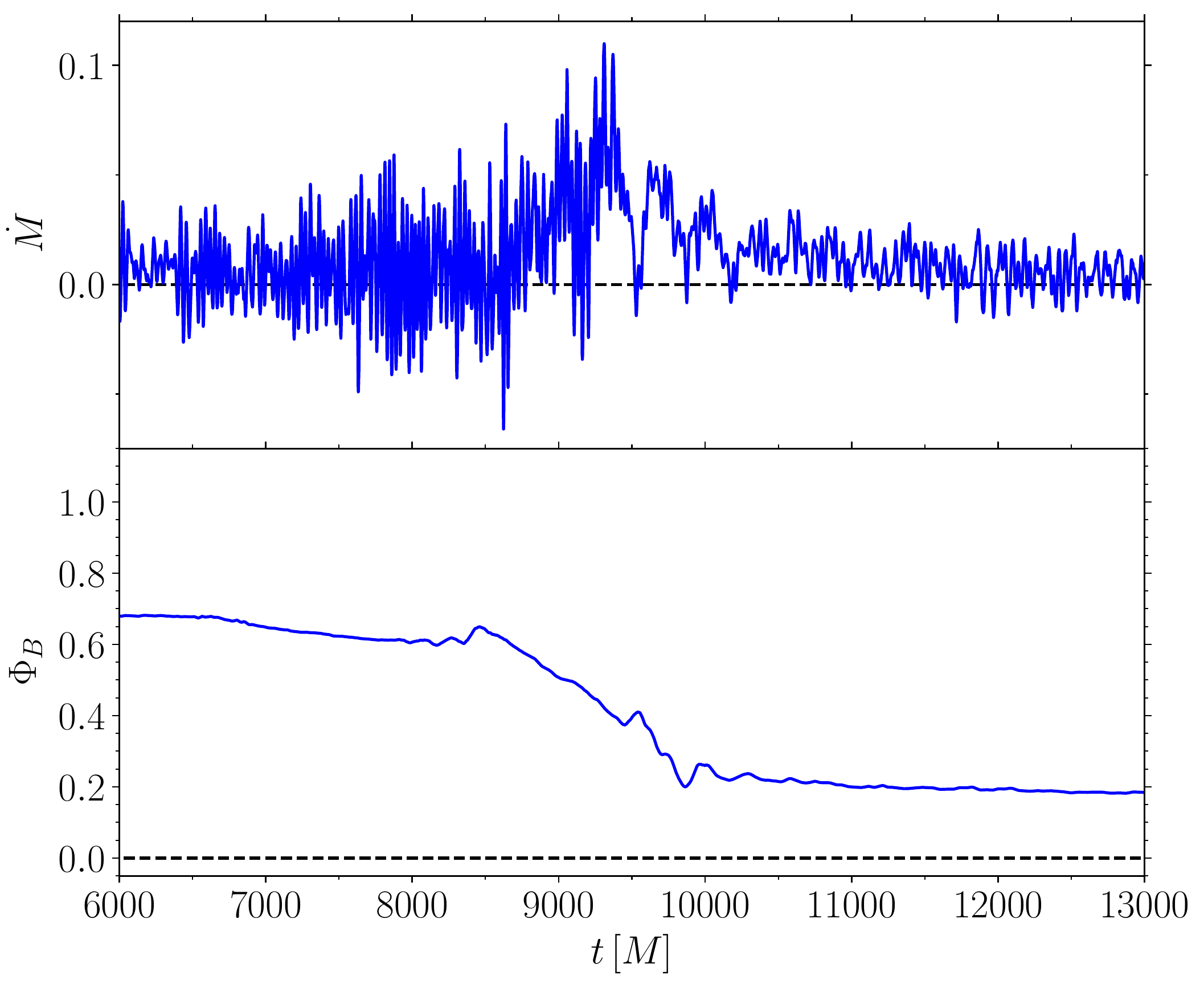}
	\caption{Same as Figure \ref{fig:GRMHD_a}, showing
		time series of the mass and absolute magnetic
		flux onto boson star model B before and after
		the changes mentioned in Appendix \ref{sec:bosonB}.}
	\label{fig:mdot_phi_bosonB}
\end{figure}

\begin{figure}
	\centering
	\includegraphics[width=\linewidth]{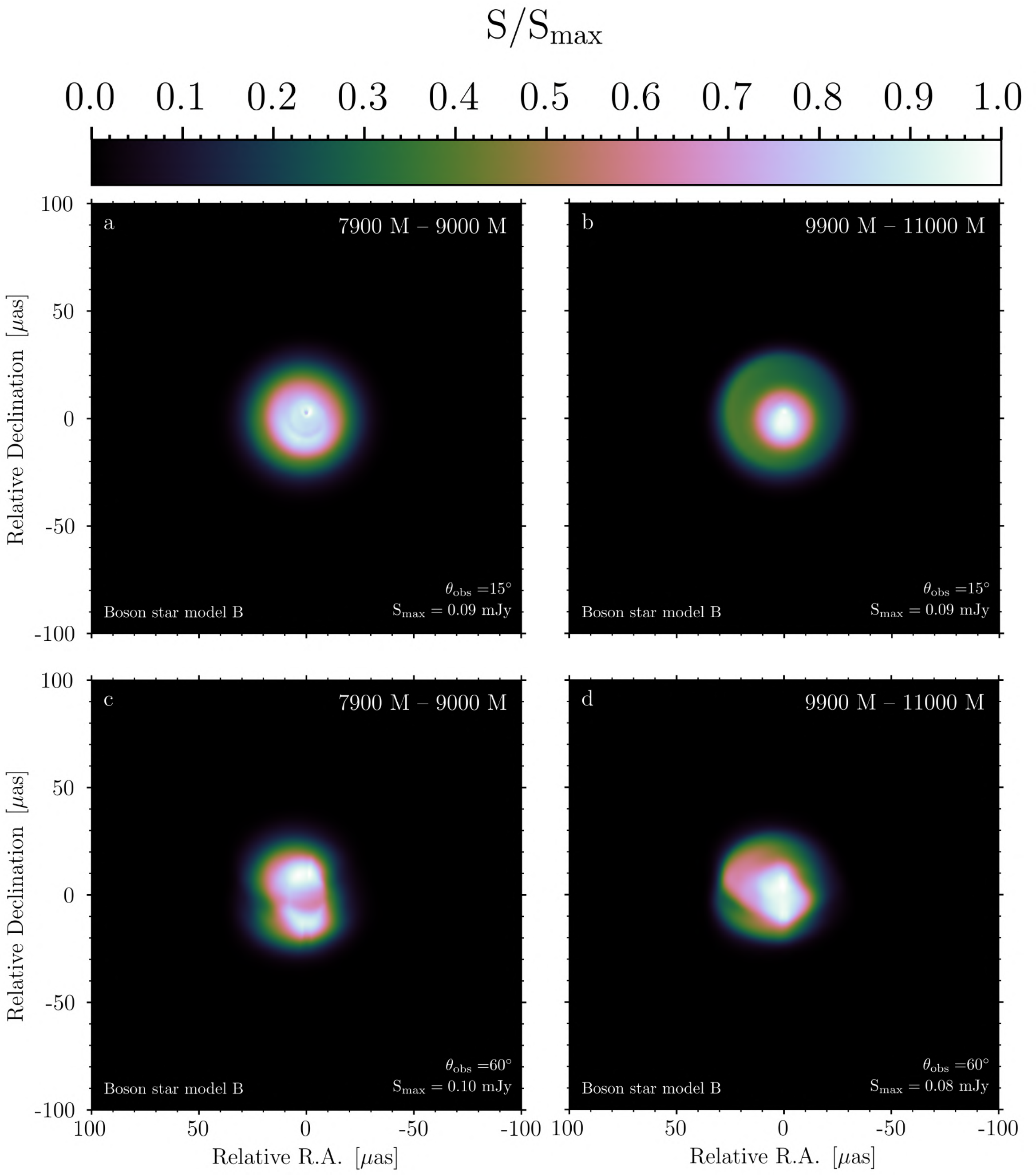}
	\caption{{ Ray-traced images at the same inclinations of Figures
		\ref{fig:grrt-images-inc15} (first row)
		and \ref{fig:grrt-images-inc60} (second row),
		averaged over the intervals $t/M\in[7900,9000]$ (left column)
		and $t/M\in[9900,11\ 000]$ (right column).}}
	\label{fig:bosonB_4panels}
\end{figure}

Between $t=8000\ M$ and $t=10\ 000\ M$, a series of changes in the
magnetic field structure produces a drop in the absolute magnetic
flux threading boson star model B (\cf Figure \ref{fig:GRMHD_a}).
These are caused by the absorption of an orbiting dense cloud by
the central fluid structure located inside the boson star.
This cloud arises from the random perturbations added to the initial
condition, and it survives and grows due to non-linear interactions
with the oscillating fluid structure inside the boson star.
In order to ensure that the images of boson star model B obtained
in the time range reported in Section \ref{sec:images} are
representative despite this changes, we ran the simulation further
until $t=13\ 000\ M$. We found that after $t=10000\ M$, the system reaches
a new long-lived state in which $\Phi_{_{B}}$ does not have rapid changes.
Figures \ref{fig:contours} and \ref{fig:mdot_phi_bosonB} show, respectively,
density isocontours and time series of the mass and magnetic flux threading
the boson star before, during and after the absorption of the cloud.
The images computed during the long-lived states before and after the
transition, and
averaged over a time window corresponding to the EHT observing time, share
the features of Figures \ref{fig:grrt-images-inc15} and
\ref{fig:grrt-images-inc60} that allow them to be distinguished from black hole
images, namely a smaller source size and the absence of a dark region
at the centre. { Figure \ref{fig:bosonB_4panels} shows images at the same
inclinations as in Figures \ref{fig:grrt-images-inc15} and
\ref{fig:grrt-images-inc60}, computed over the time windows
$t\in[7900\ M, 9000\ M]$ and
$t\in[9900\ M, 11\ 000\ M]$}, indicating that these image properties can indeed
be considered representative of this boson star model.

\subsection{Time-scale for collapse}

Since our simulations show that matter accreted onto the boson star keeps
accumulating at its interior, a natural question that arises is for how
long accretion can continue at the same rate before the accreting
material reaches the critical mass to collapse to a black hole. Although
an accurate answer to this question needs to take into account the
non-linearity of the system, a rough estimate can be made using the
accretion rate calculated in section \ref{sec:images} from the observed
flux of Sgr~A*. This calculation also ignores the effect of radiation
pressure, which could contribute to slow down the accretion flow
\citep[see][for the case of a Bondi--Hoyle--Littleton
  accretion]{Zanotti2011}. Assuming an upper limit on the accretion rate
within a 2-sphere of radius $r=2M$ that is of the order of
$10^{-6}\ {\rm M}_\odot\, {\rm yr}^{-1} \sim 10^{-12}\ M\, {\rm yr}^{-1}$,
where the second estimate refers to Sgr~A*, whose mass is $M\sim
10^{6}\,{\rm M}_{\odot}$. Considering that the scalar field in the boson stars
already accounts for $\sim 90\%$ of the mass-energy contained within a
radius $r=2M$, it would take $\sim 10^{11} {\rm yr}$ to accrete a
sufficient amount of mass to induce a collapse to a black hole.

Clearly, such a large time-scale, which corresponds to nearly seven times
the age of the Universe, suggests that the accumulation of matter in the
interior of an accreting boson star, either in the form of a mini torus
or of a mini cloud, may lead to the collapse to a black hole only on
cosmological time-scales.


\section{Initial torus and development of MRI}
\label{sec:tori}

The torus around the boson star was built according to the prescription
by \citet{Abramowicz78}, which was derived for general axisymmetric
metrics and is frequently employed for building tori around black holes.
For the boson star case, the metric functions of the Kerr space-time were
replaced by those correspondent to that of the boson star. Inside the
torus, we set up a poloidal magnetic loop from a vector potential
following density isosurfaces, $A_\phi \propto \max
((\rho/\rho_\mathrm{max} - 0.2),0)$. We adopted the following actions in
order to make the comparison between the simulated accretion flows as
close as possible:

\begin{enumerate}
\item Using the bisection method, the value of the constant angular
  momentum of the tori was set in such way that they shared the same
  inner (outer) radius of $6\,(42)\,M$.
\item We normalized the rest-mass density such that in each case it took
  the maximum value $\rho_{\rm max} = 1$.
\item We rescaled the magnetic field so that the ratio of gas to
  magnetic pressure had a minimum of $\beta_{\rm min} =11.2$.
\end{enumerate}

\begin{figure}
\begin{center}
  \includegraphics[width=0.8\linewidth]{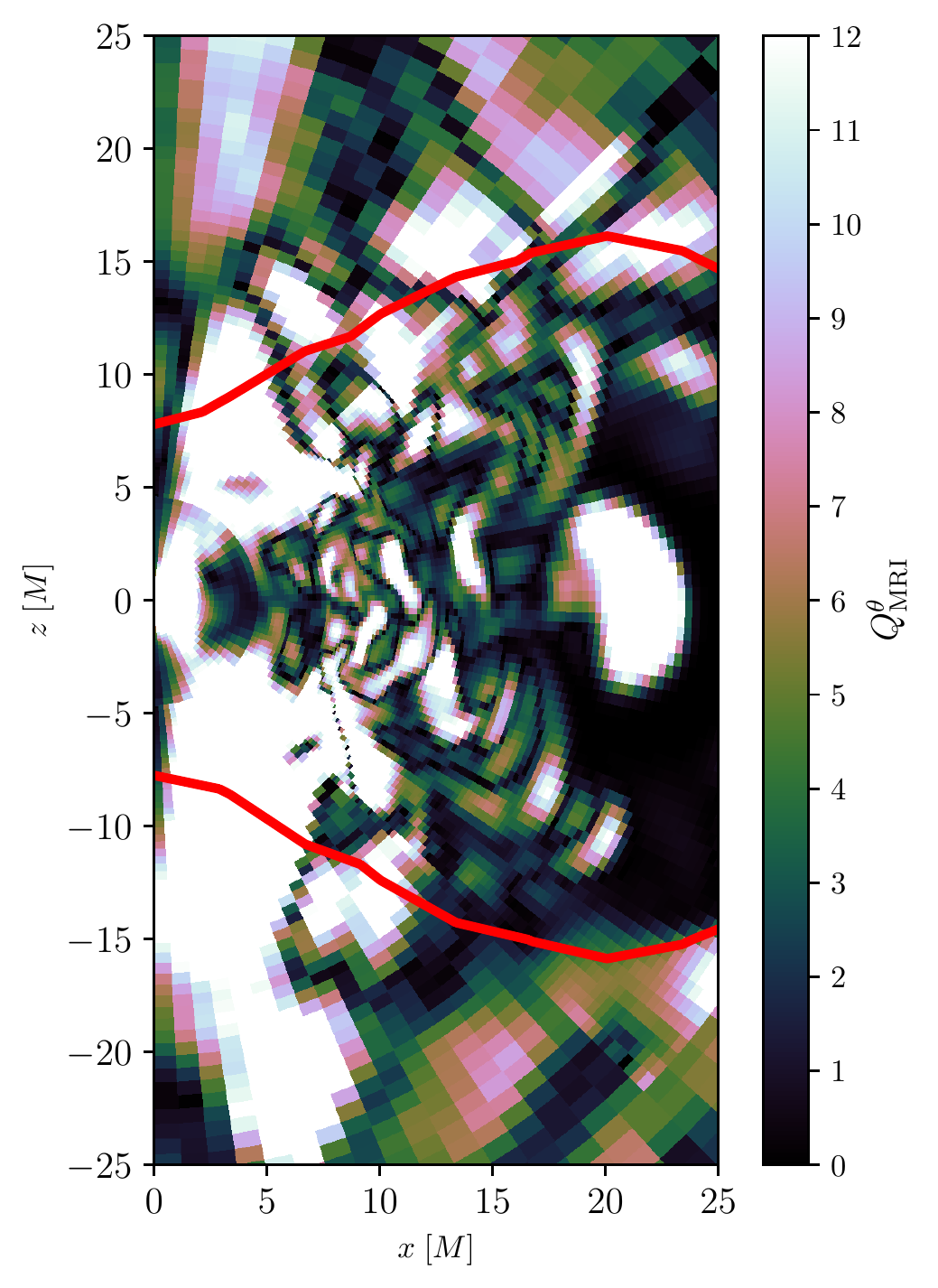}
  \caption{MRI quality factor $Q_{\rm MRI}$ for boson star model B at
    $t=1000\ M$, when the instability is developing. The red contour
    indicates the edges of the torus, approximated as the contour in
    which density falls to 1\% of the maximum density of the original
    torus. It can be seen that $Q_{\rm MRI} \gtrsim 6$ where the
    instability is developing.}
\label{fig:Q_MRI}
\end{center}
\end{figure}

The simulations were performed in polar coordinates on a grid
logarithmically spaced in the radial direction. We employed three levels
of adaptive mesh refinement triggered by the L\"ohner scheme
\citep{Loehner87}, to give an effective resolution of
$\{N_{r},N_{\theta},N_{\phi}\} \simeq \{512, 128, 128\}$, and with the
outer boundary placed at $1000\,M$, thus with a radial-grid spacing of
$0.19\,M$ at the inner edge of the torus. The accretion torus was
perturbed to trigger the MRI, causing turbulent transport of angular
momentum and driving the accretion \citep{Balbus1991}. To ensure the
ability to resolve the MRI, the resolution employed is comparable to
those encountered in the literature for simulations of accretion onto
black holes \citep[see \eg ][]{Narayan2012, Moscibrodzka2016,
  Mizuno2018}. As customary, we have computed the MRI quality factor
$Q_{\rm MRI}$ \citep[see ][]{Sano2004,Noble2010,McKinney2012}, making
sure that $Q_{\rm MRI} \gtrsim 6$ in the relevant regions
(Fig. \ref{fig:Q_MRI}), which ensures that the correct saturation values
of the shear stress and the ratio between magnetic and fluid pressure are
achieved \citep{Sano2004}.

\bsp	
\label{lastpage}
\end{document}